\documentclass[twocolumn]{aastex61}

\usepackage{graphicx}
\usepackage{xspace}
\usepackage{amsmath}
\usepackage{amssymb}
\usepackage{amsthm}

\def\eq#1{\begin{equation} #1 \end{equation}}

\def\comm#1           {{\tt (COMMENT: #1)}}
\def\sm#1             {{\tt (MACRO: #1)}}

\def\ff{$\{f_i\}$}
\newcommand{\vth}{{\boldmath{$\theta$}}}
\newcommand{\vect}[1]{\boldsymbol{#1}}

\shorttitle{Star-Galaxy Separation}
\shortauthors{Slater, Ivezi{\'c}, and Lupton}

\begin{document}

\title{Morphological Star-Galaxy Separation}

\author{Colin T. Slater}
\affiliation{Department of Astronomy, University of Washington, Box 351580, Seattle, WA 98195, USA}

\author{{\v Z}eljko Ivezi{\'c}}
\affiliation{Department of Astronomy, University of Washington, Box 351580, Seattle, WA 98195, USA}

\author{{Robert H. Lupton}}
\affiliation{Department of Astrophysical Sciences, Princeton University, Princeton, NJ 08544, USA}

\begin{abstract}
We discuss the statistical foundations of morphological star-galaxy
separation. We show that many of the star-galaxy separation metrics in common use
today (e.g. by SDSS or SExtractor) are closely related both to each other,
and to the model odds ratio derived in a Bayesian framework by \citet{Sebok}.
While the scaling of these algorithms with the noise properties of the
sources varies, these differences do not strongly differentiate their
performance.
We construct a model of the performance of a star-galaxy separator in a
realistic survey to understand the impact of observational signal-to-noise
ratio (or equivalently, $5\sigma$ limiting depth) and seeing on 
classification performance. The model quantitatively demonstrates that,
assuming realistic densities and angular sizes of stars and galaxies, 10\%
worse seeing can be compensated for by approximately 0.4 magnitudes deeper
data to achieve the same star-galaxy classification performance.
We discuss how to probabilistically combine multiple measurements, either of 
the same type (e.g., subsequent exposures), or differing types (e.g., multiple 
bandpasses), or differing methodologies (e.g., morphological and color-based 
classification).
These methods are increasingly important for observations at faint
magnitudes, where the rapidly rising number density of small galaxies makes
star-galaxy classification a challenging problem.
However, because of the significant role that the signal-to-noise ratio plays in
resolving small galaxies, surveys with large-aperture telescopes, such as
LSST, will continue to see improving star-galaxy separation as they push to
these fainter magnitudes.

\end{abstract}


\section{Introduction}

Classification of detected objects into stars and galaxies is a basic ingredient
for a wide range of science cases for astronomical surveys. At bright magnitudes
this problem is relatively simple, since stars outnumber galaxies and the
galaxies which do exist are obviously extended on the sky. As surveys push to
deeper depths, certainly in the SDSS regime and even more so with surveys on
8-meter class telescopes such as LSST, the statistics become much less favorable for
star/galaxy (hereafter abbreviated S/G) separation as galaxy numbers rapidly
increase and their apparent size on the sky decreases. This poses a considerable
risk for these surveys, particularly for Galactic and stellar science cases
where, for example, searches for rare objects or low contrast stellar
overdensities in the Galactic halo can be particularly hampered by the ``noise''
of misclassified galaxies contaminating the stellar sample.

The focus of this work is specifically on morphological separation by
distinguishing point-sources from non-point sources (other methods such as
color-based separation, e.g. \citealt{FHW}, are complementary but outside the
scope of this work). There are two main steps in
the S/G separation problem: the first is to perform some sort of measurement on
the pixel values obtained in an image with the goal of constructing a suffient statistic, and the second is to interpret these
resulting measurements as a classification into stars or galaxies (either
individually or in an ensemble).

The first automated classifiers were developed during the advent of large
digitized surveys in the 1970s, owing to the availability of high speed
microdensitometers and eventually small CCDs. This necessitated the development
of algorithms to summarize this pixel-level data. \citet{Sebok} presented one of
the first detailed analyses of the S/G separation problem in a Bayesian
framework, deriving the theoretically optimal classifier under the assumption of accurate models of the
point-spread function (PSF) and galaxies (i.e. an accurate model of the scene). \citet{Kron} developed a classifier based on the mean
value of inverse squared radius of a source, which would measure deviations from the PSF profile.

A common thread amongst many of these algorithms is the comparison of a pure
PSF-fit with a broadened measurement, where the wider profile may or may not
have free parameters. \citet{Valdes82} compared the likelihoods of the observed
source over a set of template models, which included both stellar, broadened
stellar, and measurement artifact profiles. The SDSS classifier
used the ratio of the flux in the best-fit galaxy model to the flux measured
with a pure-PSF model. As reported in \citet{Lupton2001}, ``We initially hoped
to use the relative likelihoods of the PSF and galaxy fits to separate stars
from galaxies, but found that the stellar likelihoods were tiny for bright
stars, where the photon noise in the profiles is small, due to the influence of
slight errors in modeling the PSF.'' \citet{leauthaud07} found that for
space-based data, the ratio of the peak surface brightness of an object to the
total flux performed better than the neural-network classifier used by Source
Extractor (known as \texttt{CLASS\_STAR}). More recent versions of Source
Extractor have used a parameter \texttt{spread\_model} \citep[][and also see
Section~\ref{sec:spreadmodel} herein]{Desai}, which compares the flux in the PSF
fit to the flux in a PSF broadened by a fixed factor (rather than fitting a
galaxy model). As we will argue in this work, the commonality of these methods
derives from the fact that this type of comparison is closely related to the
theoretically optimal Bayesian classification, with the primary differences
arising in the handling of noise and deviations from any simplifying
assumptions.

After one or more measurements are produced for each source in an image, the
task of assigning S/G classifications still remains. Approaches to this problem
vary considerably, ranging from the assignment of fixed cut-off values (e.g.
SDSS), to decision trees \citep{Weir95}, neural networks \citep{Odewahn1992},
hybrid ensemble methods \citep{Kim15}, or Bayesian methods incorporating priors on
the object populations \citep{henrion11}. In contrast to the similarity of most
pixel measurements, the diversity of these methods reflects the fact that there
is no single correct way to \textit{use} a classifier---for a realistic survey the
desired trade-off between completeness and contamination, and the evolution of
that desired trade-off with signal-to-noise ratio, is a choice that depends on the
specific scientific goals of the survey. The tradeoff between completeness and
purity of a sample is characterized by the receiver operating characteristic
(ROC) curves, which plots these two metrics against each other for a binary
classifier.

Our goal in this work is to provide pedagogical but fully technical answers to
the following questions:
\begin{enumerate}
  \item What is the statistical basis for the various pixel-level measurement
  techniques in common use, and how do they relate to the optimal Bayesian
  procedure outlined in \citet{Sebok}?
  \item How can the performance of a classifier on a single object be quantified
  in terms of the (candidate) galaxy size and the signal-to-noise ratio of the
  observation? In other words, what is the theoretical information content of a
  single observation of an object?
  \item For a realistic \textit{population} of objects observed in a survey,
  what overall performance can be expected under various observing conditions?
  In particular, can we expect adequate performance in case of LSST, which will
  survey the sky $\sim 5$ magnitudes deeper than SDSS? Should LSST
  use the same star-galaxy morphological classification algorithm as SDSS,  or
  could it achieve better performance?
  \item How should one combine multiple independent measurements of a star-galaxy
  separator in a statistically justifiable way? Examples could include repeated
   measurements of the same type, morphological measurements in different
   passbands, or combinations of morphological and color-based information.
    
\end{enumerate}

One topic that we will not address is the treatment of closely-spaced sources, where the light from
the sources overlaps on the image and they  cannot be treated as isolated from each other. While
blended sources are one of the principal challenges of a realistic image processing pipeline, we
wish to lay out the theory for isolated sources first without the complication of blending.

In Section~\ref{sec:optimal} we review and compare a number of classification
techniques for morphological S/G separation. In Section~\ref{sec:gaussprof} we
evaluate and compare these classifiers with simulated observations using
Gaussian light profiles. Section~\ref{sec:modeling} details realistic modeling of the
theoretically optimal performance for single objects, using populations of stars
and galaxies in a survey such as LSST. We discuss a probabilistic framework for
combining multiple independent measurements in Section~\ref{sec:probSG},
and summarize our conclusions in Section~\ref{sec:conclusions}.

\begin{figure}
  \centering
  \includegraphics{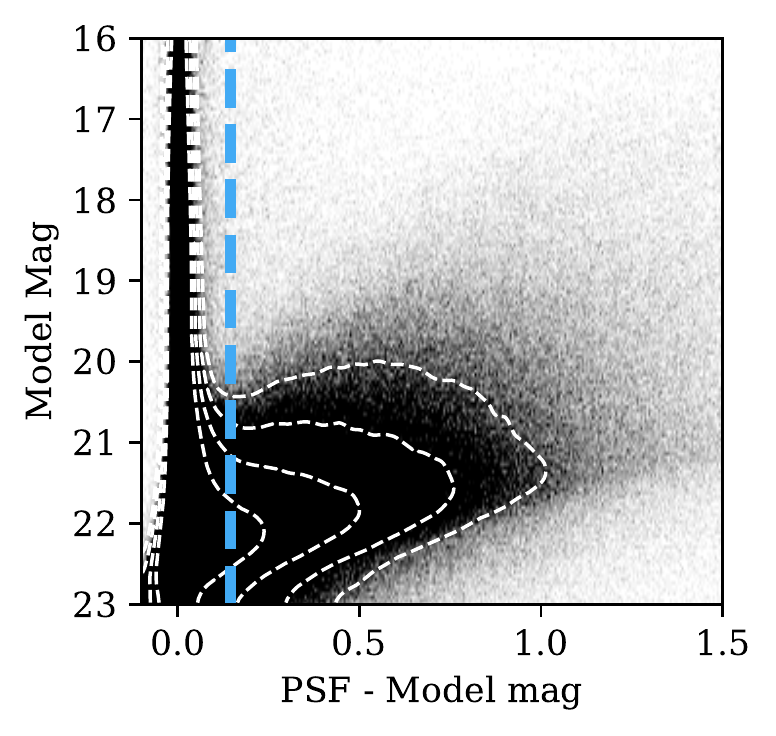}
  \caption{``PSF -- Model'' classification metric used by SDSS, shown for a set of Stripe
  82 observations. The PSF magnitude is the best fitting flux assuming the object is
  unresolved, while the galaxy model magnitude also fits for an intrinsic size of the object.
  The stellar locus is at PSF--model = 0, and objects where PSF --
  model $< 0.145$ (dashed vertical line) were classified by SDSS as stars. The
  plume of galaxies to the right of this division shows good separation at
  bright magnitudes, but at faint magnitudes tends to merge with the stellar
  locus due to both the decreasing apparent size of faint galaxies and the less
  precise measurements of faint objects. Density contours are shown as white dashed lines to illustrate the shape of the distribution.
  }
  \label{fig:Cvsmag}
\end{figure}

\section{Candidate Classifiers\label{sec:optimal}}

In this section, we derive and compare expressions for a number of metrics
derived from images that can be used for star-galaxy separation. In the
following section we compare their performance using analytic Gaussian profiles.

\subsection{Morphological Star-Galaxy Separation}

Statistically speaking, the problem of morphological star-galaxy separation
represents a case of hypothesis testing in frequentist statistics, or a case of
model selection in Bayesian statistics (for an introduction and comparison of
the two frameworks, see Chapters 4 and 5 in \citealt[][hereafter ICVG]{ICVG}).
Following the notation from ICVG, we ask whether model $G$ (galaxy) or model $S$
(star) is better supported by data $D$. Here data $D$ are represented by
measurements (counts) \ff\ for $N$ pixels, and their uncertainties, $\sigma_i$.

Models $G$ and $S$ give model predictions for data \ff. Assuming that the
profile corresponding to model $S$ is known (e.g., from analysis of bright
stars), and equal to the point-spread-function (PSF), $\phi$,
\eq{
\label{eq:fS}
  f_i^S   = C_{\textrm{psf}} \, \phi_i + {\rm noise},
}
the only free model parameter is the normalization factor $C_{psf}$, or the
so-called ``PSF'' counts (it is assumed that $\sum_{i=1}^N \phi_i = 1$). To simplify our
analysis we are assuming that the position of the source is known \textit{a priori} and is not a
model parameter. In practice, uncertainity in the source center contributes to uncertainty in S/G
separation, and covariance between the centroid and the flux measurement may need to be accounted
for in the likelihood function for accurate results.

We assume
that the galaxy model $G$ is more involved and described by a normalization
factor $C_{gal}$ (galaxy model counts) and a vector of {\it additional} free model
parameters \vth,
\eq{
\label{eq:fG}
  f_i^G   = C_\textrm{gal} \, g_i(\vect{\theta}) + {\rm noise}.
}
In the case of the SDSS galaxy models, for example, the vector of free model
parameters  has three components: axis ratio, position angle, and
characteristic radial scale, evaluated for two fixed Sersic indices (exponential
profile with $n=1$ and de Vaucouleurs profile with $n=4$) with a parameter $F_\textrm{deV}$ controling the
fractional contribution of each, and normalized such
that $\sum_{i=1}^N g_i = 1$ (again neglecting the centroid.) Our parameterization assumes that the galaxy model function
$g(\vect{\theta})$ ``knows'' the image PSF, and that the two components can be appropriately
convolved together to obtain the model pixel values. In practice, for our modeling we will assume
that the PSF and galactic light distribution are both Gaussian, and that the convolved light profile
is a Gaussian with width equal to the quadrature sum of the PSF and galaxy widths. This
assumption slightly overestimates the accuracy with which galaxies can be measured, since real
galaxies are less centrally-concentrated than a Gaussian. In this work we are most interested in the
galaxies which are only marginally resolved (and thus are closest to being distinguished by a
classifier), so the convolution of the source with the PSF dominates in setting the shape more than
the detailed galaxy model. Realistic surveys will of course fit more sophisticated models.

\subsection{The Data Likelihood}
\label{sec:datalikelihood}

A common starting point for both Bayesian analysis and the frequentist maximum likelihood
and likelihood-ratio analysis, is the likelihood of data. The data likelihood, given
a model $M=(S,G)$ and the corresponding model parameters $C$ and \vth,  as well as prior
information $I$, can be expressed as
\begin{multline}
\label{eq:dataL}
 p(D|M,C,\vect{\theta},I) = \\
   (2\pi)^{-N/2} \, \prod\limits_{i=1}^N \, \sigma_i^{-1} \,
                               \exp\left( -{(f_i - f_i^M(C, \vect{\theta}))^2 \over 2 \sigma_i^2} \right),
\end{multline}
where we assume that measurements include independent Gaussian noise parametrized by
$\sigma_i$. Although the data likelihood is often interpreted as
``the probability of the data given the model'', it is not properly normalized
to be a probability distribution function, PDF (the likelihood of 
\textit{individual} data points is a proper PDF).

In frequentist statistics, the maximum likelihood method maximizes $p(D|M,C,\vect{\theta},I)$
over model parameters $C$ and \vth\ to obtain their best-fit values, $\hat{C}$ and $\vect{\hat{\theta}}$
(note that the likelihood itself cannot be interpreted as probabilities for model parameters).
The likelihood-ratio test for two models, with likelihoods evaluated with these best-fit parameters,
is then used to select the more likely model (when competing models are not nested like here,
where the $S$ model is the same as a $G$ model with vanishing intrinsic size, various generalizations
of the likelihood-ratio test can be used instead; see \citealt{Protassov2002}). In other words, an object
is declared a galaxy when the maximum likelihood ratio,
\begin{multline}
\label{eq:Lratio}
 \Lambda \equiv {p(D|G,\hat{C}_\textrm{gal},\vect{\hat{\theta}},I) \over p(D|S,\hat{C}_\textrm{psf}, I)} = \\
  \prod\limits_{i=1}^N \,
      \exp\left(-{(f_i - \hat{C}_\textrm{gal} \, g_i(\vect{\hat{\theta}}))^2 - (f_i -
      \hat{C}_\textrm{psf} \, \phi_i)^2 \over 2 \sigma_i^2} \right),
\end{multline}
is larger than some likelihood-ratio threshold $\Lambda_{SG}$.

The assumption of Gaussianity in the second line of Equation~\ref{eq:Lratio} makes $\Lambda$ very brittle, especially in the high
signal-to-noise ratio (SNR) limit and in case of non-negligible model
errors (i.e. when observed galaxies have profiles different than those included in the model
library, or when the point spread function is not adequately modeled).

As discussed earlier, the optimal value of threshold $\Lambda_{SG}$ depends on the
desired completeness-purity tradeoff, which implies that it also reflects relative numbers of
stars and galaxies in a given sample.

\subsection{Maximum likelihood estimate for PSF counts \label{sec:MLpsfmags}}

Before proceeding with a discussion of Bayesian model selection,  we briefly review derivation of
the maximum likelihood estimate for PSF counts, $C_\textrm{psf}$. Using Equations~\ref{eq:fS} and \ref{eq:dataL},
the data likelihood is
\begin{multline}
\label{eq:psfdataL}
      p(D|S,C_\textrm{psf},I) = \\
       (2\pi)^{-N/2} \, \prod\limits_{i=1}^N \, \sigma_i^{-1} \,
                                      \exp\left( -{(f_i - C_\textrm{psf} \, \phi_i)^2 \over 2 \sigma_i^2} \right).
\end{multline}
The maximum likelihood value of $C_\textrm{psf}$, denoted as $\hat{C}_\textrm{psf}$, can be found by maximizing
the log-likelihood $\textrm{ln}\,L$:
\begin{multline}
    \textrm{ln}\, L(C_\textrm{psf}) \equiv \ln(p(D|S,C_\textrm{psf},I)) = \\
    {\rm const.} - {1 \over 2}\sum_{i=1}^N
                                            {(f_i - C_\textrm{psf} \, \phi_i)^2 \over \sigma_i^2},
\end{multline}
that is, using the condition $d(\textrm{ln}\,L)/dC_\textrm{psf}=0$. The associated uncertainty of
$\hat{C}_\textrm{psf}$ could be estimated from $\sigma_C=(d^2(\textrm{ln}\,L)/dC_\textrm{psf}^2)^{-1/2}$,
evaluated at $C_\textrm{psf} = \hat{C}_\textrm{psf}$.

Assuming homoscedastic noise, i.e., $\sigma_i \sim \sigma_0 =$ constant as is the
case when the noise is dominated by the background contribution, yields the
maximum likelihood estimate
\eq{
\label{eq:CpsfML}
               \hat{C}_\textrm{psf} = {\sum_{i=1}^N f_i \phi_i \over \sum_{i=1}^N \phi_i^2},
}
and its uncertainty (which implies a Gaussian PDF)
\eq{
\label{eq:CpsfMLsigma}
              \sigma_C  = \sigma_0  \,  \left(\sum_{i=1}^N \phi_i^2\right)^{-1/2} = \sigma_0  \,
              (n_\textrm{eff}^\textrm{psf})^{1/2}.
}
In the last expression, we have introduced the effective number of pixels, $n_\textrm{eff}^\textrm{psf}$ (the variance $\sigma_C^2$
is the sum of variances in each pixel, $\sigma_0^2$, over the effective number of pixels). For
reference, with a Gaussian psf $n_\textrm{eff} = 4\pi\alpha^2$ where $\alpha$ is the Gaussian width in
pixels, or $n_\textrm{eff} = 2.266  (\textrm{FWHM})^2$ in terms of the Gaussian full-width at half-maximum in pixels
(a three-pixel FWHM has $n_\textrm{eff} \sim 20$.)

For the case of the broadened model, the maximum likelihood model counts can be estimated as
\eq{
\label{eq:CmodML}
              \hat{C}_\textrm{gal}  = {\sum_{i=1}^N f_i g_i(\vect{\hat{\theta}}) \over \sum_{i=1}^N g_i(\vect{\hat{\theta}})^2},
}
where the vector of model parameters $\vect{\hat{\theta}}$ corresponds to the maximum likelihood point.

\subsection{The star-galaxy separation based on the likelihood-ratio test}

Using Equations~\ref{eq:Lratio}, \ref{eq:CpsfML} and \ref{eq:CmodML}, it can be shown that
\eq{
\label{eq:lnL}
 \ln(\Lambda) =  {1 \over 2\sigma_0^2} \, \left(
       \hat{C}^2_\textrm{gal} \sum_{i=1}^N g_i(\vect{\hat{\theta}})^2  -  \hat{C}^2_{psf}  \sum_{i=1}^N \phi_i^2 \right),
}
This expression can be recast as
\eq{
\label{eq:lnL2}
 \ln(\Lambda) =  {1 \over 2} \, \textrm{SNR}^2 \, \left[ \left({ n_\textrm{eff}^\textrm{psf} \over n_\textrm{eff}^\textrm{gal} }\right) \,
                                 \left({\hat{C}_\textrm{gal}  \over  \hat{C}_\textrm{psf}    }\right)^2 - 1 \right],
}
where the PSF signal-to-noise ratio is
\eq{
\label{eq:SNR}
               \textrm{SNR} = {\hat{C}_\textrm{psf}  \over \sigma_0 \, \sqrt{n_\textrm{eff}^\textrm{psf}}},
}
and the effective number of pixels for galaxy profile is
\eq{
\label{eq:neff}
         n_\textrm{eff}^\textrm{gal} =  {1 \over  \sum_{i=1}^N g_i(\vect{\hat{\theta}})^2}.
}

Equation~\ref{eq:lnL} can also be recast as
\eq{
\label{eq:chi}
           \ln(\Lambda) =  {1\over 2} \, \left( \chi^2_\textrm{psf} - \chi^2_\textrm{gal} \right) \equiv  {1\over 2} \, \Delta \chi^2,
}
where $\chi^2$ is the usual ``goodness-of-fit'' parameter, evaluated for the maximum likelihood model;
therefore, for a galaxy image larger than the PSF size $\Lambda$ increases as the point spread
function profile becomes less able to provide a good fit to the observed profile.

\subsection{SDSS Classifier}

The star-galaxy separator implemented in SDSS image processing pipeline {\it photo}
\citep{Lupton2001, Lupton2002} is equal to the difference between the point-spread-function
magnitude and the best-fit galaxy model magnitude. This magnitude difference was
named {\it concentration} by \citet{Scranton},
\begin{equation}
\label{eq:SDSSsg}
           c = m_\textrm{psf} - m_\textrm{gal}.
\end{equation}

In the nomenclature of this section, this can also be expressed as
\begin{equation}
    c_\textrm{SDSS} = 2.5 \log\left(\frac{\hat{C}_\textrm{gal}}{\hat{C}_\textrm{psf}}\right).
\end{equation}
This expression shows a strong similarity to Equation~\ref{eq:lnL2}, and indeed these
equations can be brought to a close analogy. Since the ratio
$(n_\textrm{eff}^\textrm{gal}/n_\textrm{eff}^\textrm{psf})$ increases monotonically with the ratio
$(\hat{C}_\textrm{gal}/\hat{C}_\textrm{psf})$, we can write the maximum likelihood estimate as
\eq{
\label{eq:lnL3}
       \ln(\Lambda) =  {1 \over 2} \, {\rm SNR}^2 \, \rho(\hat{C}_\textrm{gal}/\hat{C}_\textrm{psf})
       > \ln(\Lambda_\textrm{SG}),
}
where $\rho$ is some monotonic function of the $(\hat{C}_\textrm{gal}/\hat{C}_\textrm{psf})$ ratio. As a result, a source
can be classified as resolved when
\eq{
\label{eq:LambdaSG}
   \left( {\hat{C}_\textrm{gal} \over \hat{C}_\textrm{psf}} \right) > \rho^{-1} \left( {2\,
   \ln(\Lambda_\textrm{SG}) \over {\rm SNR}^2}\right) \equiv
           \left( {C_\textrm{gal} \over C_\textrm{psf}}\right)_{min}.
}
Thus for any individual object, the SDSS classifier contains the same
information as the likelihood ratio test. However, the likelihood ratio
case shows that for a range of observations, the optimal classifier must vary
with SNR. SDSS adopted a single value that was optimized for the faint end of
their data, which in practice was very effective despite not being
theoretically optimal. Because of this, it is likely that some barely-resolved
binary stars in SDSS imaging data could be recognized as such by adopting a
lower value of $({C_\textrm{gal}/C_\textrm{psf}})_\textrm{min}$ at the bright end.

\subsection{Bayesian model selection}

To find out which of the two models is better supported by data $D$, in Bayesian framework we compare
their posterior probabilities via the {\it model odds ratio} in favor of model $G$ over model $S$
\eq{
                  O_\textrm{GS} \equiv { p(G|D,I) \over p(S|D,I) },
}
where $I$ stands for ``prior information''. Note that the concept of ``model probability'' is distinctively
Bayesian. The posterior probability (a number between 0 and 1) of model $M$ ($G$ or $S$) given data $D$,
$p(M|D,I)$, follows from the Bayes theorem
\eq{
                      p(M|D,I)  =  {p(D|M,I) \, p(M|I) \over p(D|I)}.
}
The {\it marginal likelihood} or {\it evidence} for model $M$, $p(D|M, I)$, can be obtained using marginalization
(integration) over the model parameter space (spanned by $C_\textrm{gal}$ and \vth\ for model $G$, and
$C_\textrm{psf}$ for model $S$) as
\begin{multline}
\label{eq:evidence}
       E(M) \equiv p(D|M,I) = \\
       \int p(D|M,C,\vect{\theta},I) \, p(C,\vect{\theta}|M,I) \, dC \, d \vect{\theta}
\end{multline}
The evidence quantifies the
probability that the data $D$ would be observed {\it if} the model $M$ were the correct model.
The evidence is also called the {\it global likelihood} for model $M$ because it is a {\it weighted
average} of the data likelihood $p(D|M,C,\vect{\theta},I)$, with the priors for model parameters acting
as the weighting function.

The hardest term to compute is the probability of the data, $p(D|I)$, but it
cancels out when the odds ratio is considered:
\eq{
\label{eq:odds_ratio}
                 O_\textrm{GS} =  {E(G) \, p(G|I) \over E(S) \, p(S|I)} = B_{GS} \, {p(G|I) \over p(S|I)}.
}
The ratio of global likelihoods, $B_\textrm{GS}\equiv E(G)/E(S)$, is called the {\it Bayes factor}, and
is equal to
\eq{
\label{eq:Bfactor}
         B_\textrm{GS} = {\int p(D|G,C_\textrm{gal},\vect{\theta},I) \, p(C_\textrm{gal},\vect{\theta}|G,I)
         \, dC_\textrm{gal} \, d \vect{\theta} \over
                                  \int p(D|S,C_\textrm{psf},I) \, p(C_\textrm{psf}|S,I) \, d
                                  C_\textrm{psf}}.
}

The integration of the data likelihood over the model parameter space is an expensive numerical
operation. As is well known in Bayesian statistics, and first pointed out in this context by \citet{Sebok},
the variation of data likelihood around its maximum value can be approximated by a Gaussian
(unless the signal-to-noise ratio is very low). In this case, the Bayes factor reduces to the likelihood
ratio given by Equation~\ref{eq:Lratio}, with an additional term accounting for different numbers of
free model parameters. The result is related to the Bayesian Information Criterion (BIC, see e.g.,
Chapter 5 in ICVG) as
\eq{
\label{eq:BIC}
           2\,\ln(B_\textrm{GS}) \approx \Delta BIC = 2\,\ln(\Lambda) - \, M_\theta \, \ln(N),
}
where $M_\theta$ is the dimensionality of the vector of free parameters \vth, and $N$
is the number of data points. Although the maximum likelihood ratio method is now extended
with a ``penalty'' for increased number of model parameters, this approximation results in
identical classification performance as that based on $\Lambda$ alone  when the classification
cutoff is optimized rather than prescribed a priori.

In the low signal-to-noise ratio limit, integrals from Equation~\ref{eq:Bfactor} should be explicitly
evaluated. Assuming uniform priors for all model parameters,
\eq{
\label{eq:Bfactor2}
         B_\textrm{GS} = k \, {\int p(D|G,C_\textrm{gal},\vect{\theta},I) \, dC_\textrm{gal} \, d \vect{\theta} \over
                                  \int p(D|S,C_\textrm{psf},I) \, d C_\textrm{psf}},
}
where coefficient $k$ depends only on the limits for assumed priors. Here, instead of
comparing the maximum values of data likelihoods as in Equation~\ref{eq:Lratio}, the Bayes factor
now compares the {\it mean values of the two data likelihoods over the range of model parameters
allowed by priors}. Hence, the two classification methods should have different performance,
with the maximum likelihood method expected to be inferior. An example of this comparison
will be presented in the next section, including a discussion of the Occam's razor built in the
above expression.

\subsubsection{Sebok's ansatz}

\citet{Sebok} performed a similar analysis to the above in a full Bayesian framework, but added the
simplifying assumption that
\eq{
    \int p(D|S,C_\textrm{psf},I) \, p(C_\textrm{psf}|S,I) \, d C_\textrm{psf}
     = p(D | \hat{C}_\textrm{psf}, I),
}
and similarly for galaxies. This ansatz allows the integrals from Equation~\ref{eq:Bfactor} to be
replaced with a comparison of the maximum likelihood flux estimates for both models, and results in
a likelihood calculation of the same form as Equation~\ref{eq:lnL}. \citet{Sebok} then
required that the term in square brackets in Equation~\ref{eq:lnL2} be positive to classify
an object as galaxy. This requirement yields a condition
\eq{
\label{eq:SebokC}
           \left({\hat{C}_\textrm{gal}   \over   \hat{C}_\textrm{psf}}\right) \,  \left({
               n_\textrm{eff}^\textrm{psf} \over n_\textrm{eff}^\textrm{gal} }\right) ^{1/2}  > 1.
}
Compared to Equation~\ref{eq:lnL2}, Sebok's ansatz and the assumption of equal priors on star and
galaxy density results in the dependence on SNR vanishing.

\citet{Sebok} also simplifies the evaluation of this condition by using a single fixed-size
galaxy model for comparision, on the basis that it is primarily the marginally-resolved galaxies
where the classification is most sensitive and the PSF size dominates in those cases.
Despite these differences, we show in the Section~\ref{sec:gaussprof} that the condition given by Equation~\ref{eq:SebokC} acts
similarly to the SDSS galaxy separator (in case of Gaussian profiles, Equations~\ref{gaussCrat} and \ref{gaussNrat}
imply a monotonic relationship between the two separators).

\subsection{New Star-Galaxy Separator in SExtractor}
\label{sec:spreadmodel}

We also consider a parameter called \texttt{spread\_model}, computed by the code
SExtractor \citep{Bertin}, that has been recently developed as part of the Dark
Energy Survey Data Management program \citep{DESDM}.
According to \citet{Desai}, \texttt{spread\_model} is a superior star-galaxy classification
parameter compared to \texttt{class\_star}, SExtractor's traditional star-galaxy separator
(see their Figure 13). The distribution of sources in the \texttt{spread\_model}  vs.
apparent magnitude diagram is reminiscent of the $m_\textrm{psf} - m_\textrm{gal}$ vs. magnitude diagram
constructed with SDSS data (Figure~\ref{fig:Cvsmag}).

The \texttt{spread\_model} parameter is a normalized simplified linear discriminant between
the best-fitting local PSF model ($\Phi$) and a slightly more extended model ($G$) made
from the same PSF convolved with a circular exponential disk model with scale length equal
to FWHM/16 (here FWHM is the full width at half-maximum of the PSF model). It is defined
as \citep{Desai}
\begin{equation}
\label{eq:Cspread}
          \texttt{spread\_model} =  { G^T x \over G^T \Phi} - {\Phi^T x \over \Phi^T \Phi}
\end{equation}
where $x$ is the image vector centered on the source; see also \citet{Soumagnac}. The
corresponding expression in \cite{Desai} has a sign error, which we corrected above (E. Bertin, priv. comm.).
For $x=\Phi$, $\texttt{spread\_model} = 0$ by construction, and for resolved sources,
$\texttt{spread\_model} > 0$.

SExtractor also computes \texttt{spreaderr\_model}, the uncertainty for \texttt{spread\_model} parameter.
Using this uncertainty, \citet{Bechtol} compute weighted mean of \texttt{spread\_model} for a set
of images with varying depth, and \citet{Koposov} propose a criterion for binary star-galaxy separation
that accounts for deteriorating signal-to-noise ratio close to the faint end
\begin{equation}
    | \texttt{spread\_model} | < 0.003 + \texttt{spreaderr\_model}.
\end{equation}

Both $G$ and $\Phi$ in Equation~\ref{eq:Cspread} are normalized to the observed source flux (in the
maximum likelihood sense, c.f. \S~\ref{sec:MLpsfmags}). It is easy to show,  using nomenclature
from this section, that
\begin{equation}
\label{eq:Cspread2}
          \texttt{spread\_model} = \eta \, \left({\hat{C}_\textrm{gal}  \over  \hat{C}_\textrm{psf}
          }\right)  \left({ n_\textrm{eff}^\textrm{psf} \over n_\textrm{eff}^\textrm{gal} }\right)^{1/2}  - 1,
\end{equation}
where
\begin{equation}
               \eta \equiv  {(\sum_{i=1}^N \phi_i^2)^{1/2} \, (\sum_{i=1}^N g_i^2)^{1/2} \over \sum_{i=1}^N \phi_i g_i},
\end{equation}
and with an important caveat that the model profile $g$ is fixed, rather than optimized. For
a given seeing profile, $\eta$ is fixed and, with the chosen $g$, very close to unity (to within
a few percent, see Equation~\ref{gaussEtarat} below). Hence,
as a comparison with Equations~\ref{eq:lnL2} and \ref{eq:SebokC} reveals, \texttt{spread\_model} parameter
is essentially equivalent to the classifier proposed by Sebok (apart from the fact that here $g$ is fixed).

\subsection{Summary of different classifiers}
\label{sec:Csummary}

As shown above, there are five closely related candidate classification parameters:
\begin{enumerate}
\item SDSS classifer, $C_\textrm{SDSS} = 2.5\log\left({\hat{C}_\textrm{gal}  \over  \hat{C}_\textrm{psf}    }\right)$.
\item From Equation~\ref{eq:SebokC}, \\ $C_\textrm{Sebok} = \left({\hat{C}_\textrm{gal}  \over \hat{C}_\textrm{psf}    }\right)
                             \left({ n_\textrm{eff}^\textrm{psf} \over n_\textrm{eff}^\textrm{gal} }\right)^{1/2}  \\
                             = 10^{(0.4\,C_\textrm{SDSS})} \, \left({ n_\textrm{eff}^\textrm{psf}
                             \over n_\textrm{eff}^\textrm{gal} }\right)^{1/2}$.
\item From Equation~\ref{eq:Cspread2}, $C_\textrm{spread} = \eta \, C_\textrm{Sebok} -1$.
\item From Equation~\ref{eq:lnL2},  \\ $C_{\Delta \chi^2} = \left( \chi^2_\textrm{psf} -
    \chi^2_\textrm{gal} \right) = SNR^2 \, \left( C_\textrm{Sebok}^2 -1\right)$.
\item From Equation~\ref{eq:Bfactor}, $C_\textrm{Bayes} = 2\,\ln(B_\textrm{GS})$. \\ For high SNR,
    $C_\textrm{Bayes}  \approx C_{\Delta \chi^2}  - M_{\theta} \, \ln(N)$.
\end{enumerate}
In addition, when analyzing the behavior of Gaussian profiles in the next
section, we will also consider the best-fit profile width, $C_\sigma$, as the sixth
classification parameter.

Note that the first three classification parameters do not include dependence on the signal-to-noise
ratio SNR. In the high SNR limit, all six classifiers are expected to have similar performance.
Our aim in the next section is to quantify their behavior in the low SNR limit, using Gaussian profiles.
On general statistical grounds, we expect $C_\textrm{Bayes}$ to perform the best, and seek to quantify whether
its performance gain compared to, e.g., $C_\textrm{SDSS}$ or $C_{\Delta \chi^2}$, might be significant in practice.

\section{Comparison of Classifiers in case of Gaussian profiles \label{sec:gaussprof}}

\subsection{Outline}

In order to compare the statistical properties of the six classifiers summarized in the preceding
section, we use an idealized case based on Gaussian profiles for both the source and the PSF.
The main goal is to compare their performance in the low-SNR limit. Given SNR, specified by
the total number of counts and (Gaussian) noise per pixel, and the values of the PSF width,
$\theta_\textrm{psf}$, and the intrinsic source width, $\theta_g$, we generate a large number of sources
(10,000) and equal number of the corresponding PSFs. The profile variations are entirely due to
random realizations of the assumed noise. For each source, we fit two free parameters, the profile
width and its normalization, using uniform priors (that is, the best fit corresponds to maximum
likelihood solution). Given these best-fits, we evaluate the six classification parameters (note
that for $C_\textrm{Bayes}$ fitting can be bypassed) and compare their distributions for the sources
and for the PSFs. Instead of specifying a priori classification threshold, we evaluate the full
ROC (receiver operating characteristic) curve, which is a standard tool for quantifying the
completeness vs. contamination tradeoff. Although for some classifiers there are some
pre-defined numerical values, e.g. a threshold value of 1 for $C_\textrm{Sebok}$, or the properties of
fixed galaxy profile in case of $C_{spread}$, we optimize over them for a fair comparison of all
classifiers. We study the performance of classification parameters as a function of SNR and
 the $({\theta_g / \theta_\textrm{psf}})$ ratio. In the rest of this section, we discuss and
illustrate these steps in more detail.

\subsection{Model profiles and fitting method}

We assume a circular Gaussian profile
\eq{
\label{eqn:singleG}
         p(r|\alpha) = {1 \over 2 \pi \alpha^2}  \exp{\left(-{r^2 \over 2 \alpha^2}\right)},
}
which satisfies $2\pi \int_0^\infty p(r|\alpha) r dr = 1$, and $\theta \equiv$ FWHM=2.355$\alpha$.
The counts from a source are then described by
\eq{
\label{eqn:sourceProf}
                       C(r) = C \, p(r|\alpha) + n(\sigma_0),
}
where $n(\sigma)$ is Gaussian noise with a mean of zero and standard deviation equal to
$\sigma$. The source profile width parameter $\alpha$ is a result of the convolution
of the PSF and an intrinsic source profile, and is obtained by
\eq{
         \alpha = (\alpha_\textrm{psf}^2 + \alpha_g^2)^{1/2}.
}
We use $\alpha_\textrm{psf}$=1.5 pixel, corresponding to $\theta_\textrm{psf}=3.5$ pixel
(motivated by the median expected seeing for LSST, which has the same $\theta_\textrm{psf}$
is pixel units). Given this large $\theta_\textrm{psf}$, for simplicity we evaluate the profile at the
pixel center.
For a Gaussian profile, $n_\textrm{eff} = 4\pi(\alpha/{\rm pix})^2 = 28.3$.

\begin{figure*}
  \centering
  \includegraphics{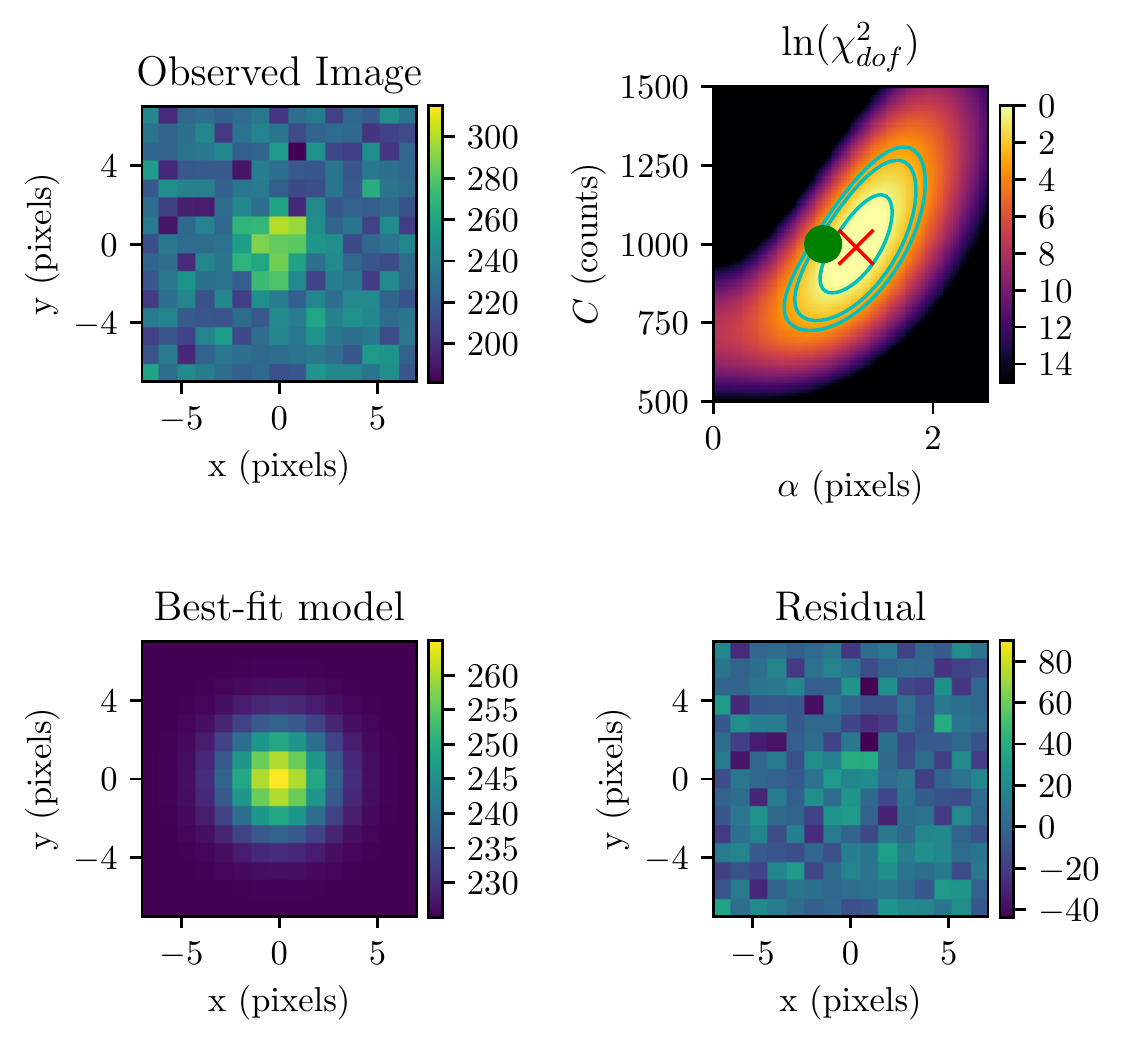}
  \caption{
    An illustration of fitting an image generated with noise per pixel of $\sigma_0=15$ counts,
     PSF with $\alpha_\textrm{psf} =1.5$ pix, the intrinsic profile width $\alpha_{g} =1.0$ pix, and
     a source with $C=1000$ counts. With an effective number of pixels of $\sim$40,
     the SNR is $\sim$10. The top left panel shows data image and the top right panel
     shows the $\chi^2$ image as a function of two free parameters, $\alpha_{g}$ and $C$.
     The standard $1\sigma$, $2\sigma$ and $3\sigma$ contours are shown by the lines,
     the maximum-likelihood best-fit values of the free parameters by the $\times$ symbol, and
     the input values of fitted parameters by the dot. The best-fit model is shown in the
     bottom left panel, and the data-model residuals in the bottom right panel.
  }
  \label{fig:lnL_bestfit}
\end{figure*}

We get the best-fit values of $C$ and $\alpha_g$ by a grid search. Given a
15 pix by 15 pix image generated with chosen input values of $C$, $\alpha_\textrm{psf}$,
$\alpha_g$ and $\sigma_0$, we compute the data likelihood $L$ using Equation~\ref{eq:dataL}
as a function of two free model parameters, $C$ and $\alpha_g$ (and a related quantity
$\chi^2 = -2\ln(L)$). The maximum likelihood best-fit is a pair of ($C$, $\alpha_g$)
values that maximizes $L$ (or minimizes $\chi^2$). An example of such fitting is shown
in Figure~\ref{fig:lnL_bestfit}.

\subsection{Analytic predictions for Gaussian profiles \label{sec:anProfs}}

Before proceeding with numerical experiments, we summarize analytic predictions for the behavior
of classifiers for the case of noise-free Gaussian profiles. For profiles described by
$\alpha_\textrm{psf}$
(PSF) and $\alpha_g$ (source; before convolution with the PSF), it can be shown analytically (and
numerically in case of $\eta$) that in the noise-free case,
\eq{
\label{gaussCrat}
    \left({C_\textrm{gal}  \over  C_\textrm{psf}    }\right) = 1 + {1 \over 2} \, \left({\alpha_g
    \over \alpha_\textrm{psf}}\right)^2,
}
\eq{
 \label{gaussNrat}
         \left({ n_\textrm{eff}^\textrm{gal} \over n_\textrm{eff}^\textrm{psf} }\right)  = 1 +
         \left({\alpha_g \over \alpha_\textrm{psf}}\right)^2,
}
and
\eq{
 \label{gaussEtarat}
        \eta = 1 + 0.06 \, \left({\alpha_g \over \alpha_\textrm{psf}}\right)^3.
}

Hence, all classifiers are only functions of the ratio $(\alpha_g/\alpha_\textrm{psf})$ and are {\it uniquely}
related to each other. {\it The differences in the statistical behavior of classifiers are due only to their
varying response to noise,} as quantitatively discussed below.

The above expressions also elucidate the behavior of classifiers that explicitly depend on SNR.
For example, it follows from Equation~\ref{eq:lnL2} that ,
\begin{equation}
 C_{\Delta \chi^2} = {\rm SNR}^2 \, \left( {C_\textrm{gal}^2 \, n_\textrm{eff}^\textrm{psf} \over
    C_\textrm{psf}^2 \, n_\textrm{eff}^\textrm{gal}} -1 \right) =
                         {\rm SNR}^2 \, \rho\left(\alpha_g/\alpha_\textrm{psf}\right),
\end{equation}
where $\rho(\alpha_g/\alpha_\textrm{psf})$ is a monotonic function of the ratio
$(\alpha_g/\alpha_\textrm{psf})$ (and note a close relationship to Equation~\ref{eq:lnL3}).
This expression shows that, given a threshold for $C_{\Delta \chi^2}$, the smaller values of $\alpha_g$
can be ``resolved'' at a higher SNR. \textit{When the star-galaxy separation threshold is optimized at
the faint, low-SNR, end of the data, the ability to recognize barely resolved objects at the bright
end is not fully exploited}. This behavior is generic and not limited to Gaussian profiles as long
as $\rho$ is a well-defined monotonic function of $(\alpha_g/\alpha_\textrm{psf})$.

\subsection{Illustration of the $\chi^2$ behavior in low SNR case}

\begin{figure*}
  \begin{center}
    \plottwo{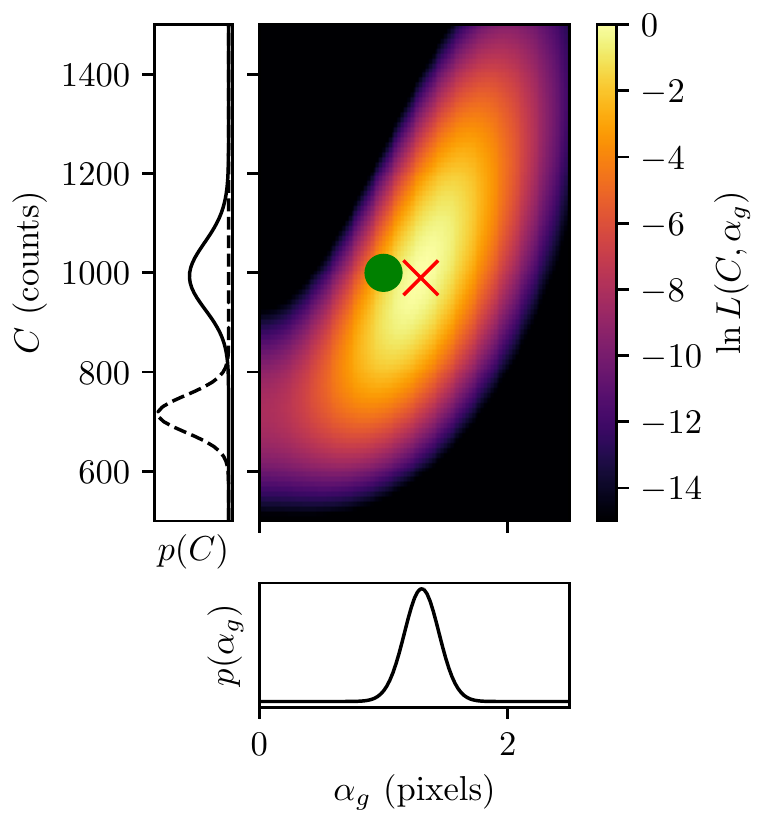}{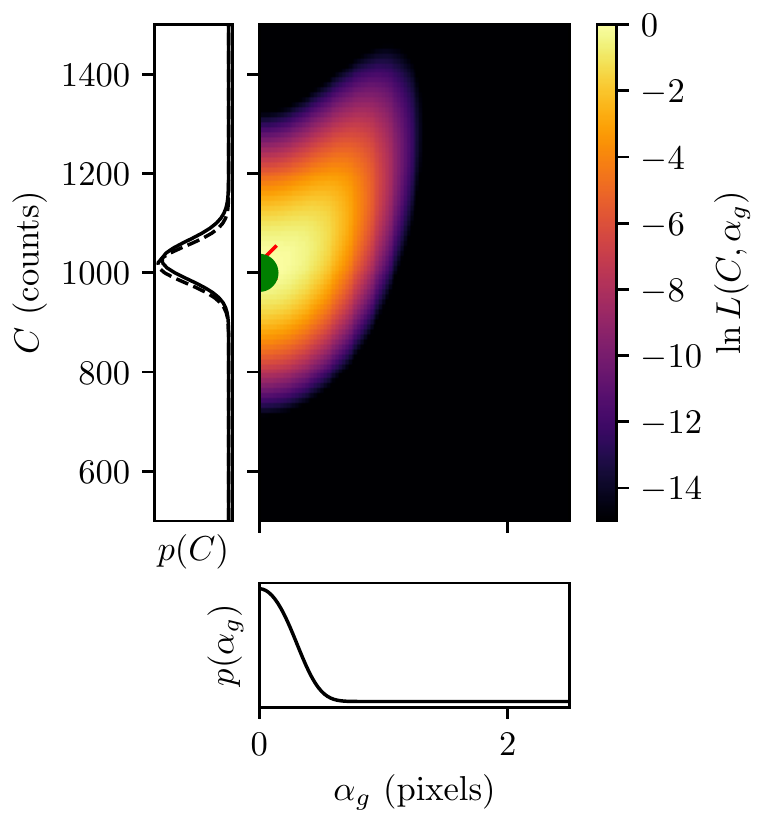}
 \end{center}
  \caption{
The left panel shows the two-dimensional log-likelihood surface (ln(L)=$-\chi^2/2$) for fitting
the intrinsic profile width ($\alpha_g$) and normalization ($C$, see Equation~\ref{eqn:sourceProf})
of an image generated with noise per pixel of $\sigma_0=15$ counts, PSF width $\alpha_\textrm{psf} =1.5$ pixels,
and $\alpha_{g} =1.0$ pixels (same image as shown in the top left panel in Figure~\ref{fig:lnL_bestfit}).
The circle marks the true values, and the $\times$ symbol marks the maximum likelihood point. Marginal
probability distributions for each parameter are shown to the left and below the panel with solid
lines. The dashed line in the panel to the left is the conditional distribution of the normalization
$C$ given $\alpha_g=0$ (note that its peak corresponds to the maximum likelihood value of
PSF counts, $C_\textrm{psf}$). The right panel is analogous, except for a profile with $\alpha_{g} =0$
(a noisy realization of the PSF profile). Note that the marginal distributions for $\alpha_{g}$
deviate from a Gaussian shape, especially in the right panel.
}
\label{fig:lnL}
\end{figure*}

In the high-SNR regime, the likelihood surface around the maximum likelihood point, such
as that shown in the top right panel in Figure~\ref{fig:lnL_bestfit}, can be well approximated
by an elliptical Gaussian. As the SNR decreases, the deviations from Gaussianity can be large,
with details depending on the profile parameters. Examples of low-SNR likelihood surfaces
are shown in Figure~\ref{fig:lnL}.

\subsection{Comparison of Different Classifiers}

\begin{figure*}
  \vskip -0.01in
  \centering
  \includegraphics{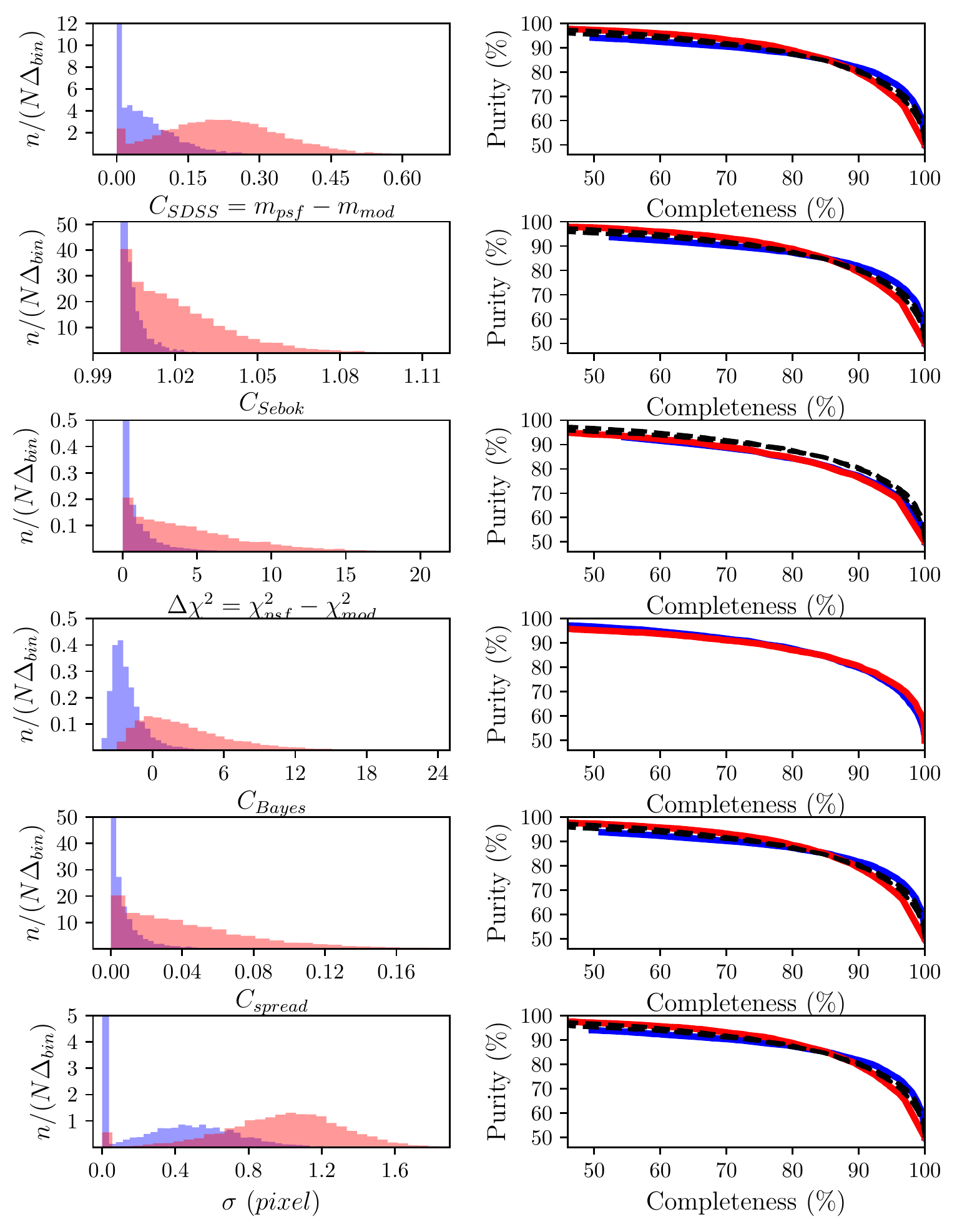}
  \caption{
    A comparison of the performance of six different classifiers (see \S\ref{sec:Csummary} for definitions)
    for simulated data. The left column shows distributions of classifier values for 10,000 realizations of a star or galaxy,
    with noise per pixel of $\sigma_0=15$ counts, with blue histograms
    corresponding to PSF-like sources ($\textrm{SNR} = 12.5$)
    and red histograms to a profile with $\alpha_{g} =1.0$ pix and the same total source counts
    ($\textrm{SNR} = 10.4$.)
    The right column shows purity vs. completeness ROC curves stars (blue line)
    and galaxies (red line), where the number of true stars and galaxies in the
    sample are equal.  The dashed black line is the stellar ROC curve for
    $C_\textrm{Bayes}$, replicated to the other plots to aid comparison.
  }
  \label{fig:Class1_14}
\end{figure*}

\begin{figure*}
  \centering
  \includegraphics{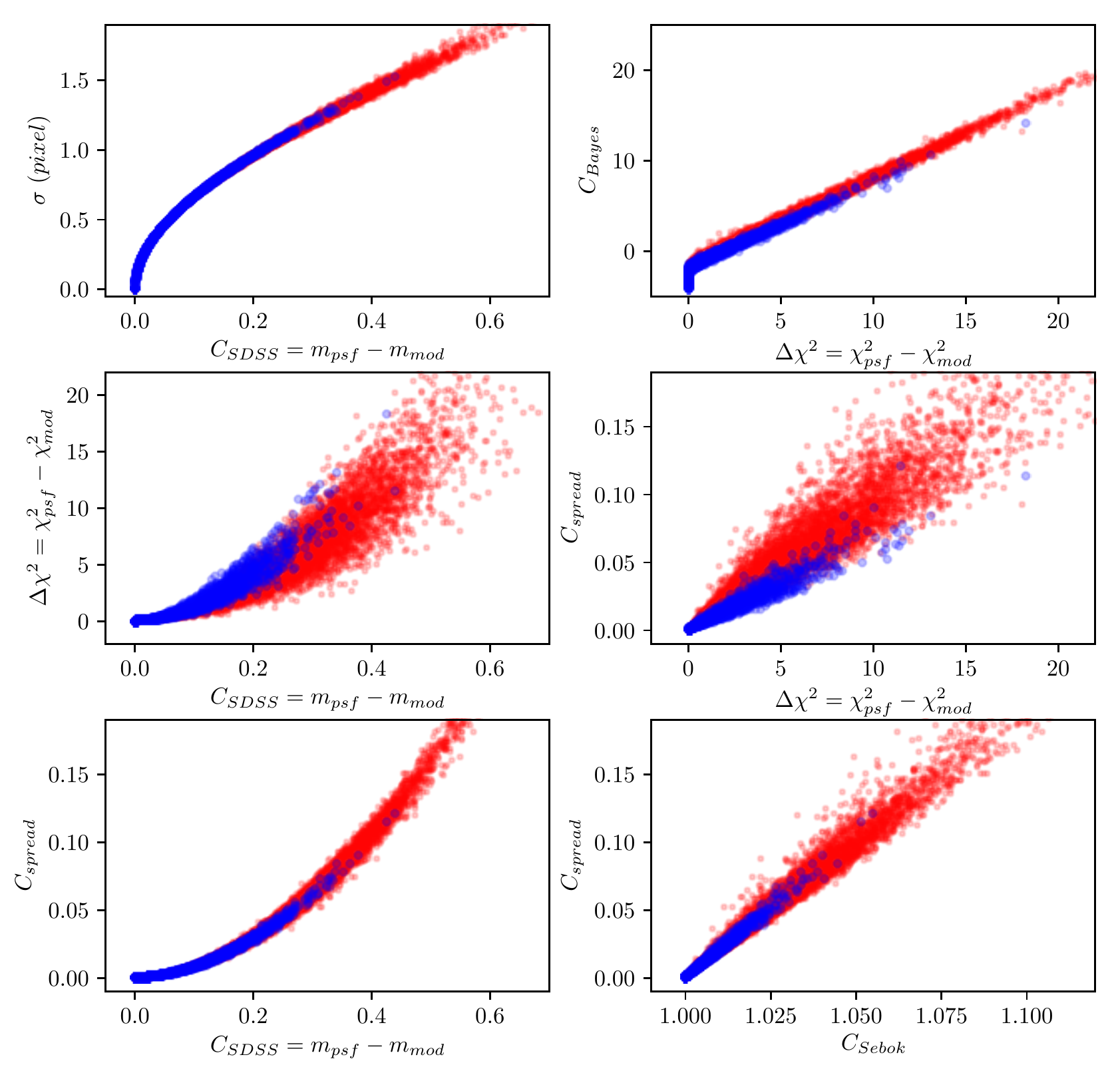}
  \caption{
    An illustration of correlations between six different classifiers (see \S\ref{sec:Csummary} for definitions).
    The values for 10,000 random draws are shown, with noise per pixel of $\sigma_0=15$ counts,
     with blue symbols corresponding to PSF-like sources ($\alpha_\textrm{psf} =1.5$ pix,
     $\alpha_{g} =0$) and red symbols to a
     profile with the same PSF and $\alpha_{g} =1.0$ pix. The distributions of classifier values are shown in
     Figure~\ref{fig:Class1_14}.
  }
  \label{fig:Comp1_14}
\end{figure*}

To compare the behavior of the different classifiers to each other, we created
10,000 random ``star'' images (with $\alpha_g=0$ and $\alpha_\textrm{psf} = 1.5$) and
10,000 ``galaxy'' images ($\alpha_g = 1.0$), all at $\rm{SNR} \approx 10$, which were then
classified by all of the algorithms we have considered. Classifier histograms
and ROC curves for each classifier's values are shown in
Figure~\ref{fig:Class1_14}, while the different classifiers are plotted against
each other in Figure~\ref{fig:Comp1_14}. In the ROC panels of
Figure~\ref{fig:Class1_14}, the $C_\textrm{Bayes}$ curve is shown in each panel by the
black dashed line, to provide a visual reference between panels. It is clear
from these curves that none of the classifiers do better than the Bayesian
result, and with possibly the exception of the $\chi^2$ classifier, none of them
do substantially worse either. This is despite the rather varied appearance of
the classifier histograms; these differences in the values returned for each
object do not translate into improved S/G performance, as shown by the ROC
curves.

Since we used analytic (Gaussian) profiles, this behavior is easy to understand
quantitatively. Expressions listed in \S\ref{sec:Csummary} and \S\ref{sec:anProfs}
imply that all classifiers are functions of the $(\alpha_g / \alpha_\textrm{psf})$ ratio, and
thus are uniquely (though non-linearly) related to each other (e.g., $C_\textrm{SDSS}$ as
a function of the best-fit profile width, see top left panel in Figure~\ref{fig:Comp1_14}).
The scatter in expected one-to-one relations is seen when at least one of the
two plotted classifiers includes explicit SNR dependence (which varies around
the input value due to random noise). For example, using expressions listed in
\S\ref{sec:Csummary} and \S\ref{sec:anProfs}, it is straighforward to show that
for Gaussian profiles (see  middle left panel in Figure~\ref{fig:Comp1_14})
\begin{equation}
 C_{\Delta \chi^2} = {\rm SNR}^2 \, { \left( 10^{0.4C_\textrm{SDSS}}  - 1 \right)^2 \over
    2\,10^{0.4C_\textrm{SDSS}}  - 1   }.
\end{equation}

\begin{figure*}
  \vskip -0.01in
  \centering
  \includegraphics[width=0.49\textwidth]{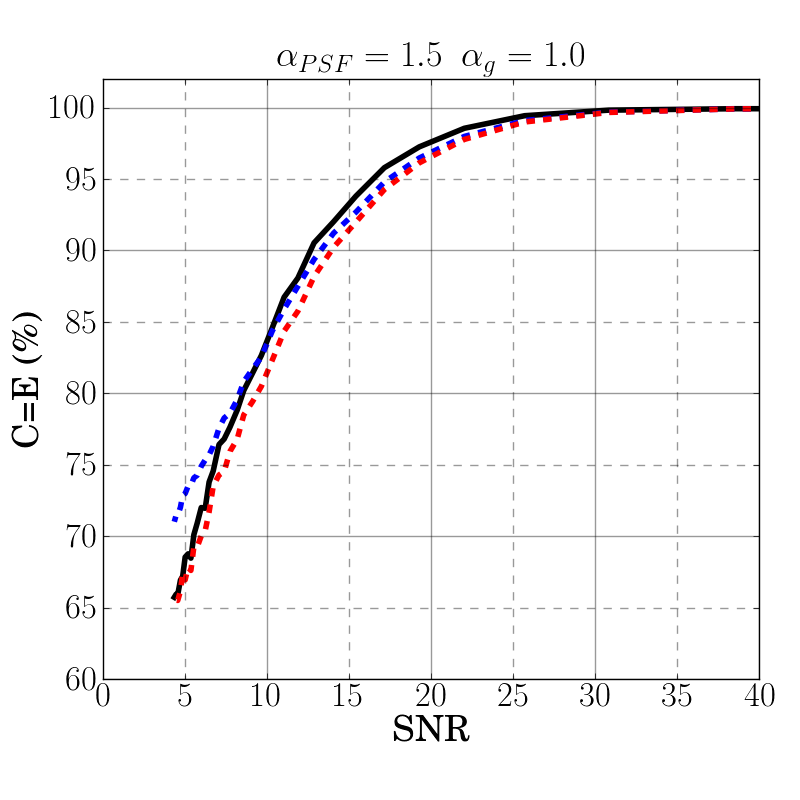}
  \includegraphics[width=0.49\textwidth]{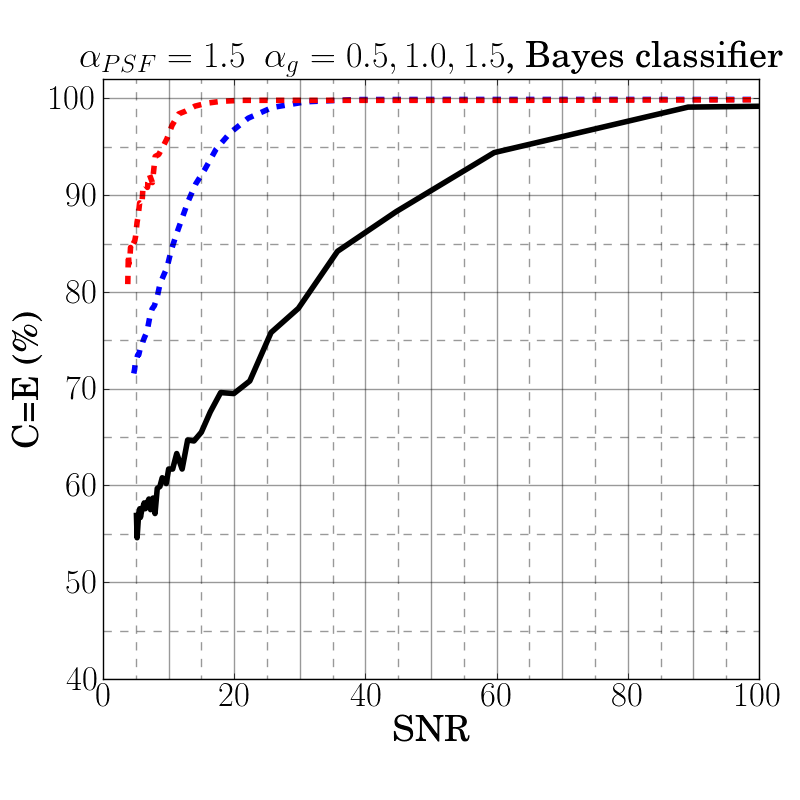}
  \vskip -0.01in
  \caption{
    The left panel compares classification performance of different classifiers using
    the Completeness = Purity point on the ROC curves as metric, for the case
    $\alpha_\textrm{psf}=1.5$ pix and $\alpha_g=1.0$ pix. The blue dashed
    line corresponds to $C_\textrm{Bayes}$, the red dashed line to $C_{\Delta \chi^2}$ and the
    other four classifiers are indistinguishable from the black solid line. Note that at
    SNR=5,  $C_\textrm{Bayes}$ exceeds the performance of other classifiers by about 7\%.
    The right panel compares the performance of $C_\textrm{Bayes}$ for three different
    values of $\alpha_g$ (solid: 0.5 pix; blue dashed: 1.0 pix; red dashed: 1.5 pix).
  }
  \label{fig:CEvsSNR1}
\end{figure*}

The left panel of Figure~\ref{fig:CEvsSNR1} shows the performance of these
classifiers as a function of observation SNR. To do so, it is necessary to
distill each ROC curve into a single number; we do so by somewhat arbitrarily
reporting the value at which the completeness is equal to the purity. While this
is not necessarily the same choice that one would make when performing S/G
separation, it is proportional to the overall performance of the classifier (a
different choice of a fiducial point will not change the result). While the
Bayesian classifier performs somewhat better at very low SNR,
at moderate and high SNR the behavior of all classifiers is very similar.

Figure~\ref{fig:CEvsSNR1} also shows how the Bayesian classifier performs for
galaxies of different sizes and at different SNR, but with fixed PSF size. For galaxies
similar in size to the PSF (blue dashed line), the performance rapidly improves
from relatively modest increases in SNR. Similar gains are also available for
objects significantly smaller than the PSF, but only at significantly higher SNR.

We reiterate that Bayesian classifier, although statistically optimal in cases when
profile models are known, is brittle in practice and very sensitive to deviations
of the observed profiles from assumed models (e.g., galaxies with dust lanes or
tidal tails, stars at high SNR).

\section{Modeling Star-Galaxy Separation Performance}
\label{sec:modeling}

In this section we show how to predict the star-galaxy separation
performance of a particular set of observations, starting only from a
pixel-level statistical model of the measurement of individual objects and a
simple model of the population of stars and galaxies in the target field. The
goal of this modeling is to extract the dependence of star-galaxy separation on the
seeing and the signal-to-noise ratio of a given set of observations, and also to
show how high SNR observations are able to resolve galaxies with angular size smaller than
the size of the PSF.

We first describe how to compute the covariance matrix for objects of a given
size and flux using the Fisher matrix formalism. We then combine this
result with the true population of stars and galaxies in a given field to obtain
a distribution of classifier values as it would be measured in those
observations. This is effectively a convolution of the underlying distribution,
with the convolution kernel varying across the size-magnitude plane.

\subsection{Modeling individual objects}

For modeling S/G performance, we need to predict the uncertainty distribution of
a chosen star-galaxy separation metric for an object of a given size---which
could be zero in the case of a star---and magnitude, under any set of observing
conditions. While the Monte Carlo approach from the previous section could be
extended to this use case, it is both more illustrative and computationally
tractable to approach this analytically using toy models. As we have shown, the various candidate
classifiers are in general more similar than they are different, so we will
focus on measuring the width of a Gaussian PSF, which may be broadened if the
object is a galaxy. We assume that the galaxy light profile is also intrinsically
Gaussian. Because the galaxies which are on the verge of being resolved will
still have their shape dominated by the PSF, the detailed light distribution in
the galaxy is a secondary effect in this context.

We can compute the minimum variance that an unbiased estimator of the observed
object size $\hat{\alpha}$ would have via the Fisher information matrix
$\mathcal{I}_{\vect{\theta}}$. This is defined as
\begin{equation}
  \mathcal{I}_{\vect{\theta}} = \mathbb{E}_{I_1, ...,I_n\sim f^n_{\vect{\theta}}}
  \Big[ \left(\frac{d}{d\vect{\theta}} \ln L(I_1,...,I_n; \vect{\theta})\right)^2\Big]
\end{equation}
where $I_n$ is the measurement of the $n$-th pixel, $\vect{\theta}$ is a vector of parameters, and $\mathbb{E}$
denotes the expectation value with respect to the function $f_{\vect{\theta}}$ from which pixel values are drawn.

Our treatment of the Fisher information follows that of \citet{mendez13}, which
showed how to compute the Cram\'{e}r-Rao bound on astrometric measurements.
In this case our interest will be in the accuracy of object size measurements,
though the methodology is largely similar.
By assuming that the likelihood function can be separated into the
product of the likelihoods for each pixel in an object (that is, the noise is not correlated
between pixels), one can show that the Fisher
information for the measurement of model parameter $\theta$ is
\begin{equation}
    \mathcal{I}_\theta = \sum_{i=0}^{N}
    \frac{1}{\sigma_i^2}
    \left(
    \frac{\partial F_i(\theta)}{\partial \theta}
    \right)^2,
    \label{eq:fisherItheta}
\end{equation}
where $F_i(\theta)$ denotes the expected value for pixel $i$ of the model being
fit to the observations, that is, the noise-free version of
Equation~\ref{eq:fG}. The full derivation of this equation is presented in
Appendix~\ref{appendix}. That derivation assumes a Gaussian noise
distribution on each pixel, but the result is the same for Poisson noise.

For star-galaxy separation, we will evaluate the Fisher
information for the measurement of both the Gaussian width ($\alpha$) and the total flux of
the object ($C_\textrm{gal}$) under assumed model
\begin{equation}
  F_i(C_\textrm{gal}, \alpha) = C_\textrm{gal} \, g(r_i, \alpha) + B,
\end{equation}
where $g(r_i, \alpha)$ denotes the value of a unit-normalized Gaussian of width
$\alpha$, evaluated at the radius $r_i$ of pixel $i$, the total flux is given by
$C_\textrm{gal}$, and the background flux $B$.

Evaluating Equation~\ref{eq:fisherItheta} with this model, we obtain
\begin{equation}
  \mathcal{I}_\alpha = \sum_{i=1}^{N} \sigma^{-2}_i \left(C_\textrm{gal} \frac{dg(r_i, \alpha)}{d\alpha}\right)^2
  \label{eqn:I_alpha}
\end{equation}
\begin{equation}
    \mathcal{I}_{\alpha \textrm{Cmod}} =   \sum_{i=1}^{N} \sigma^{-2}_i g(r_i, \alpha) \frac{dg(r_i, \alpha)}{d\alpha}
\end{equation}
\begin{equation}
  \mathcal{I}_\textrm{Cmod} =   \sum_{i=1}^{N} \sigma^{-2}_i g(r_i, \alpha)^2.
\end{equation}
The Fisher matrix containing these elements must then be inverted to obtain the
covariance matrix.

In the modeling that follows, we evaluate these functions numerically to compute
the S/G performance. To guide our understanding of these results, however, we
will also derive a simplified version that characterizes the overall behavior.

Our main quantity of interest is $\mathcal{I}_\alpha$, which is inversely
proportional to the variance in estimates of the object size. Inserting the
Gaussian derivatives into Equation~\ref{eqn:I_alpha} produces
\begin{equation}
  \mathcal{I}_\alpha = \sum_{i=1}^{N} \frac{C_\textrm{gal}^2 \left(\frac{r_i^2}{\alpha^3} -
    \frac{2}{\alpha}\right)^2 g(r_i, \alpha)^2}{C_\textrm{gal} \, g(r_i,\alpha) + B}.
\end{equation}
For background-limited observations, we can make the simplifying assumption that
individual pixels in the object have values significantly less than the
background, such that $C_\textrm{gal} \,g(r_i,\alpha) < B$. This produces
\begin{equation}
  \mathcal{I}_\alpha = \sum_{i=1}^{N} \left(\frac{r_i^2}{\alpha^3} - \frac{2}{\alpha}\right)^2
  \frac{C_\textrm{gal}^2 \, g(r_i,\alpha)^2 }{B}.
\end{equation}
This expression can be simplified further by incorporating
Equations~\ref{eq:SNR} and \ref{eq:neff}, and noting that that the $r_i^3 /
\alpha^3$ term is only significant compared to the $2/\alpha$ term at large
radii, but these radii have low weighting in the summation because of the
Gaussian function $g(r_i, \alpha)$. The resulting simplified version is
\begin{equation}
    \mathcal{I}_\alpha \sim \frac{(\textrm{SNR})^2}{\alpha_\textrm{psf}^2 + \alpha_{g}^2}.
\end{equation}

Seeing and signal-to-noise ratio thus have similar effects on the ability
to resolve galaxies. But while seeing has traditionally been understood as the
key variable in S/G separation, in practice typical seeing for modern ground-based
observatories rarely varies by more than a factor of three to four, even between different
sites. The signal-to-noise ratio is far less constrained, and for objects of a
given brightness can grow by significant factors either through longer exposure
times, larger telescopes, or reduced background. This makes the signal-to-noise
ratio the key factor to consider when planning observations or analyzing S/G
separation results.

With this qualitative understanding in hand, we can now proceed to a
quantitative estimation of S/G performance.

\subsection{Modeling Object Populations}

We model here distributions of two populations, stars and galaxies, as functions
of flux (magnitude) and size.

\subsubsection{Galaxy Size and Magnitude Distribution}
\label{sec:COSMOS}

The general behavior of the underlying distribution of galaxies in the size-magnitude
plane is that galaxies become smaller and significantly more numerous at fainter magnitudes.
The quantitative description of this distribution is best extracted from space-based data,
where there is little confusion between stars and galaxies at the angular sizes that we are
concerned with in ground-based observations. For this purpose we use a model that was developed for
LSST performance optimization studies, which models the galaxy size distribution as a log-normal function.
This was fit to the HST COSMOS-based mock catalogs of \citet{jouvel09}, which were constructed for
optimization of dark energy experiments.

\begin{figure}
  \includegraphics{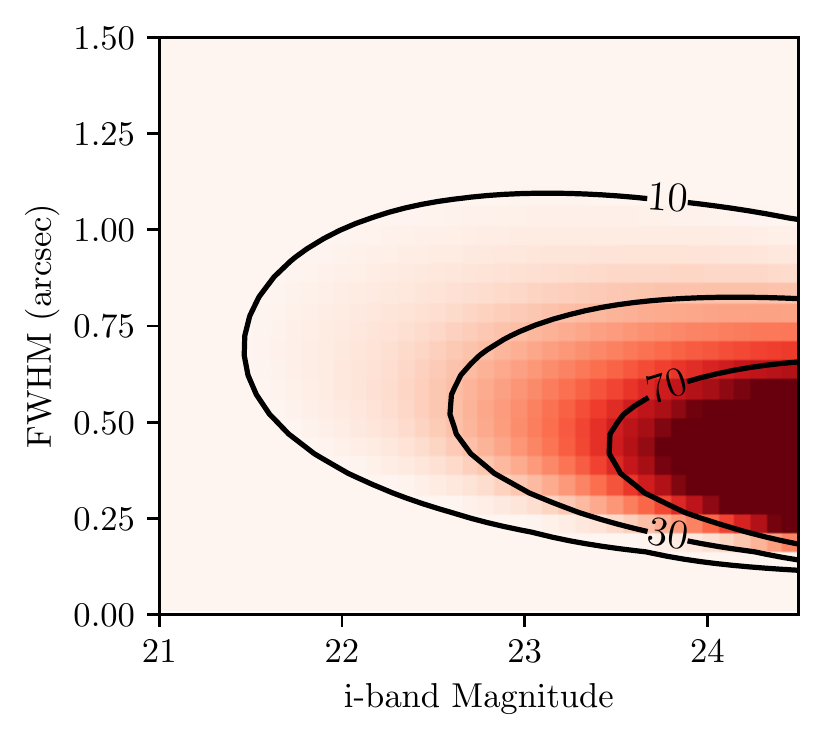}
  \caption{Model for the distribution of galaxies in the size-magnitude
  plane, as described in Section~\ref{sec:COSMOS}. The
  model assumes a log-normal distribution in size and an exponential
  distribution over magnitude. The contours are labeled in thousands of stars per square degree per
   arcsecond FWHM per magnitude.
  \label{fig:fisher_plots1}}
\end{figure}

\begin{figure}
  \centering
  \includegraphics{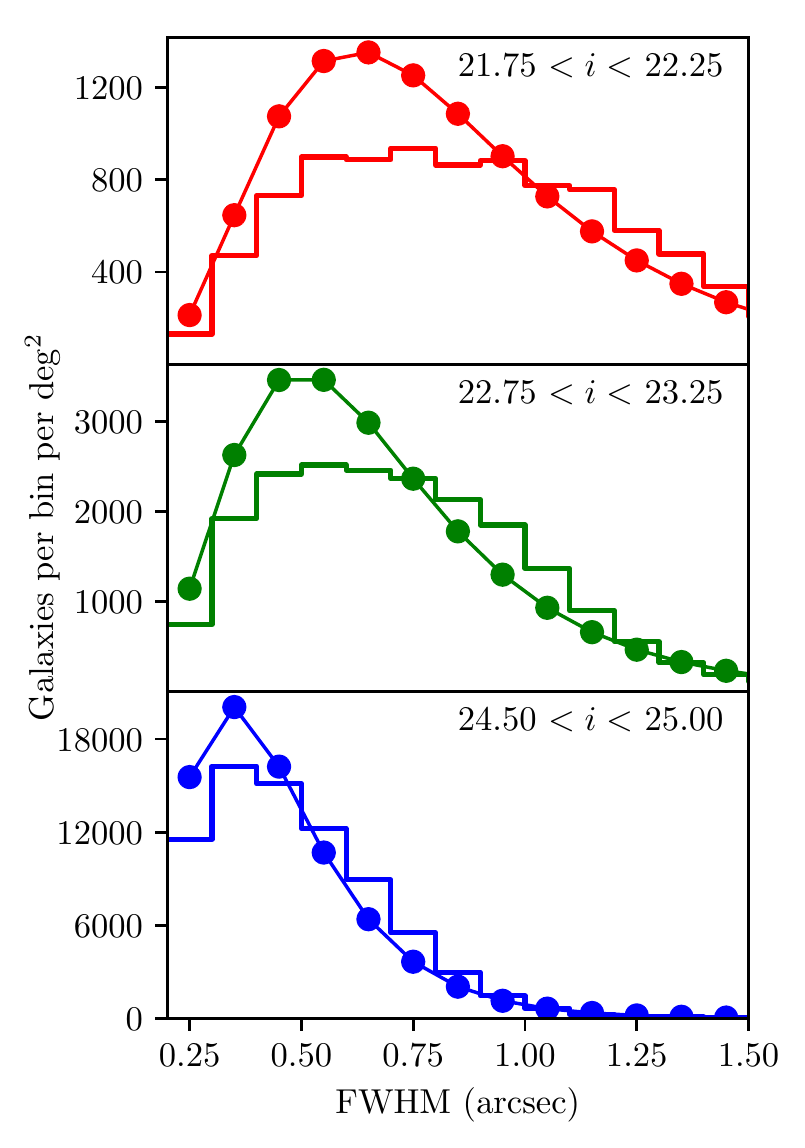}
    \caption{
        Comparison of the galaxy size distribution from our model (points connected by lines) and
        the measured object sizes from COSMOS (stepped histogram), in three i-band magnitude bins.
        While the log-normal model is somewhat more peaked and the real distribution slightly
        flatter, the overal trend of size and the number counts as a function of magnitude are
        sufficiently similar for our modeling needs.
        \label{fig:cosmos_size_magnitude}
  }
\end{figure}

The size
distribution as a function of magnitude is well described by a log-normal
distribution,
\eq{
\label{eq:logNorm}
  p(x=\theta_\textrm{gal}|\mu_\theta,\sigma_\theta)= (x\sigma_\theta\sqrt{2\pi})^{-1}\,
 \exp\left({-\frac{(\ln x - \mu_\theta)^2}{2\sigma_\theta^2}} \right),
}
whose parameters are linear functions of the $i$-band magnitude:
\eq{
    \mu_\theta = -0.24 \, i + 5.02 \,\, \ln(\textrm{arcsec})
}
and
\eq{
         \sigma_\theta = -0.0136\, i + 0.778\,\, \ln(\textrm{arcsec}).
}
The median intrinsic galaxy size (FWHM, in arcsec) is equal to exp($\mu_\theta$) and
it varies from $\sim1.0$ arcsec at $i=21$ to 0.35 arcsec at $i=25.3$. An
illustration of this model is shown in Figure~\ref{fig:fisher_plots1}.
Figure~\ref{fig:cosmos_size_magnitude} shows a comparison between these model predictions and the
observed COSMOS catalogs from \citet{capak07}. The observed sizes show a somewhat broader
distribution than the log-normal model, but the trends with magnitude are overall sufficiently
representative for our modeling.

For modeling the density of galaxies on the sky, we use a fit to data from the CFHTLS
Deep survey \citep{hoekstra06,gwyn08} from the LSST Science Book \citep{science_book}.
The cumulative galaxy counts between $20.5 < i < 25.5$ are given by
\begin{equation}
  N_\textrm{gal} = 45 \times 10^{0.31(i - 25)} \,\, {\rm arcmin}^{-2}.
\end{equation}

\subsubsection{Stellar Density Distribution}
\label{sec:stellar_density}

In addition to the model of galaxy counts, our model must also incorporate a
stellar density distribution as a function of apparent magnitude. This choice of
distribution is subject to much greater variation than the galaxy
distribution, as the position of any given survey pointing relative to the Milky
Way disk has a very significant impact on the overall normalization and shape of
the observed distribution.

Additionally, studies that target stellar samples
(and for whom galaxies are the contaminant) are in general not interested in
\texttt{any} star, they are nearly always tuned to select stars with specific
properties that make them tracers of some target phenomenon. For example,
studies of the Milky Way stellar halo often select main sequence turn-off (MSTO) stars
via their color because they have roughly constant luminosity and can be used
to probe varying distance intervals. The apparent magnitude distribution of MSTO
stars thus acts primarily as a proxy for the density distribution as a function of
distance and it essentially reflects the structure  of the Milky Way. In contrast, the
apparent magnitude distribution of a sample of stars without this color-based
selection tends to show increasing numbers of red, low-luminosity disk stars at
faint magnitudes, even at high Galactic latitudes, and thus also reflects
the steepness of the main sequence luminosity function.

\begin{figure}
  \begin{center}
  \includegraphics{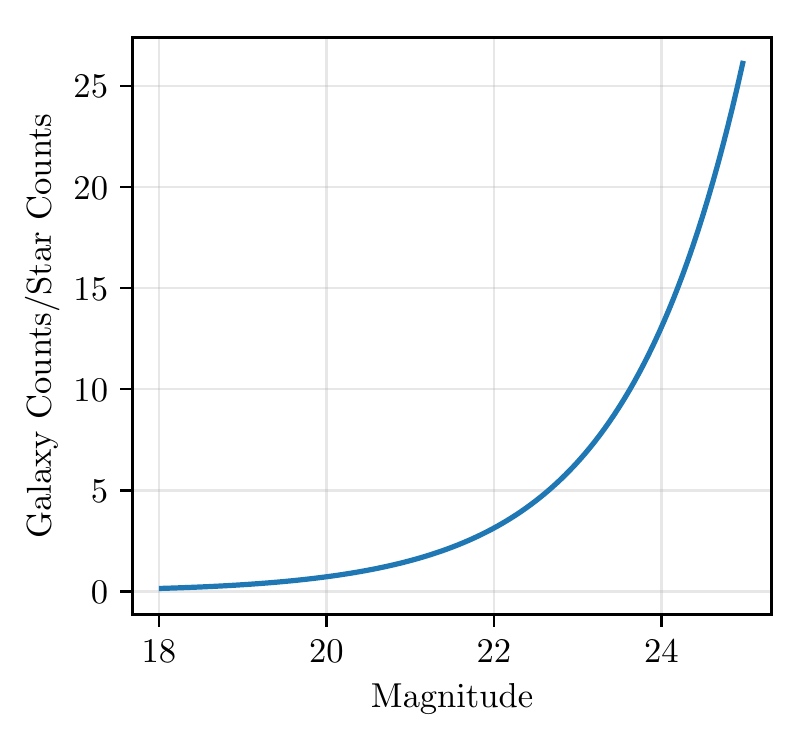}
  \end{center}
  \caption{Ratio of the galaxy density to stellar density in the input model. The galaxy and stellar densities are equal at magnitude 20.5.
  \label{fig:model_predictions_sg_ratio}}
\end{figure}

In our modeling we will assume that the target stellar sample is the Milky
Way halo, rather than nearby stars in the disk. The density profile of the halo
can be modeled as a power law which transitions from $r^{-2.5}$
inside of $\sim 25$ kpc \citep{juric08} to approximately $r^{-3.6} \sim r^{-4}$
in the outer halo \citep{slater16, cohen17}. Rather than creating a detailed
model for the density distribution along a particular line of sight through the
halo, we adopt a constant stellar number density per unit magnitude, which
corresponds to an $r^{-3}$ density profile. This is meant to provide a broadly
representative approximation of the true halo density profile. As discussed
above, the normalization of this profile is strongly dependent on Galactic
latitude and longitude, along with color-based selection criteria. We chose to
normalize the stellar density such that it equals the density of galaxies at
magnitude 20.5, and emphasize that because of the uncertainties in this choice,
our focus will be on the variation in S/G separation performance rather than any
absolute statements about completeness or purity in a given scenario.
Figure~\ref{fig:model_predictions_sg_ratio} shows the ratio of stars to galaxies in our
model as a function of magnitude.

\subsubsection{Convolving the Galaxy Distribution}

\begin{figure*}
  \begin{center}
    \includegraphics{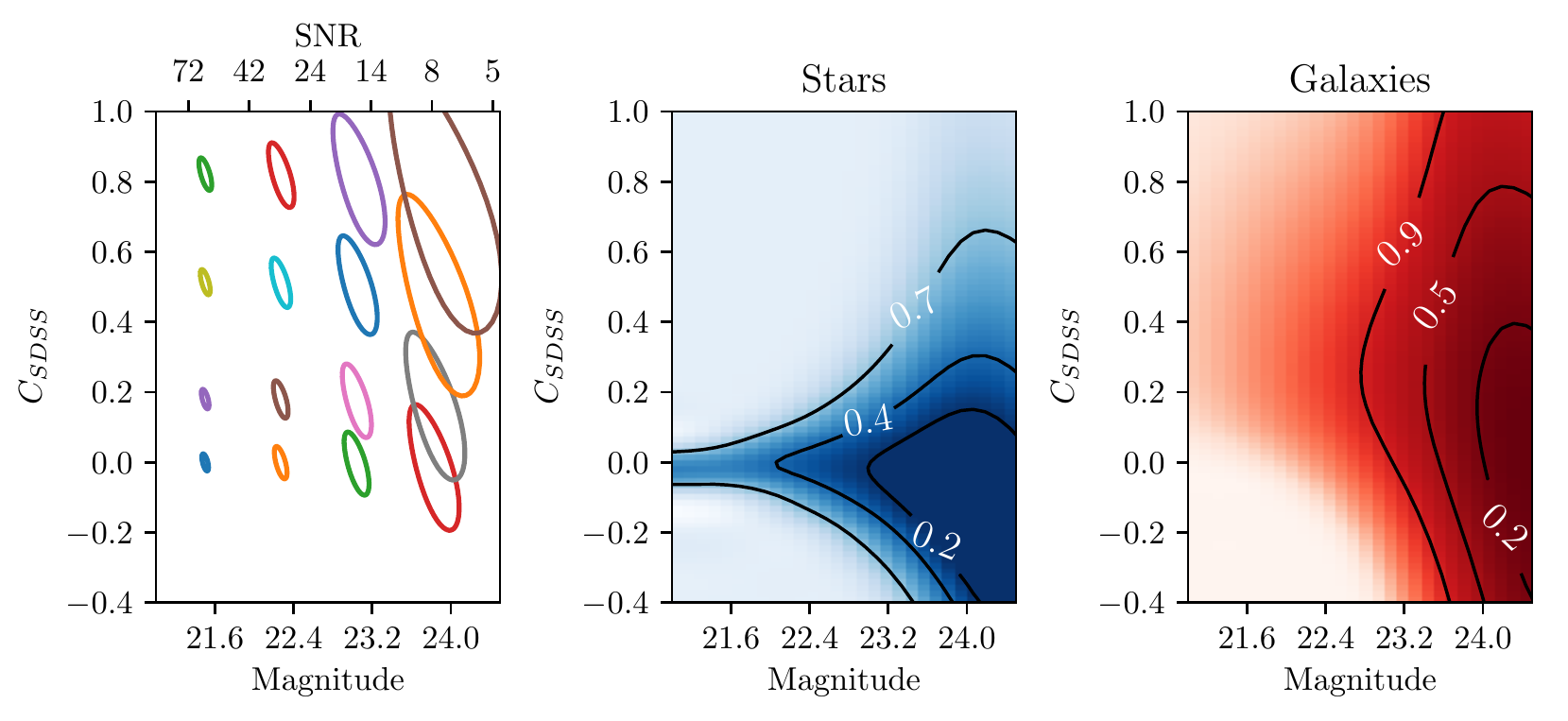}
  \end{center}
  \caption{An illustration of the modeling procedure described in
  Section~\ref{sec:modeling}.  The left panel shows the covariance ellipses for observations
  of a grid of galaxies; galaxies at faint magnitudes (or alternatively, low
  SNR) have large uncertainties in the S/G separation parameter $C_\textrm{SDSS}$,
  while high SNR objects are tightly constrained. The middle panel shows the
  distribution of measured stellar parameters (magnitude and $C_\textrm{SDSS}$)
  after convolution by the covariance ellipses, while the right panel shows the
  same model output for galaxies. The contours are labeled with the fraction of objects falling inside each
  contour. The stars intrinsically lie along $C_\textrm{SDSS} =
  0$, while the galaxies are intrinsically distributed over a wide range of
  widths (as shown in Figure~\ref{fig:fisher_plots1}).
  \label{fig:fisher_plots2}}
\end{figure*}

The left panel of Figure~\ref{fig:fisher_plots2}, shows examples of the
covariance contours from the Fisher matrix modeling that we use as convolution
kernels. On a dense grid covering the size-magnitude space, we evaluate for each
grid cell the number of galaxies that would be intrinsically present in that
cell, then compute the covariance matrix and thus the contribution of that cell
to each cell in the as-observed size-magnitude distribution. Summing over these
cells produces the distributions shown in the center and right panels of
Figure~\ref{fig:fisher_plots2}. The trend for the ``observed'' distribution of galaxies
to be both wider (due to increased counts and observational scatter) and shifted
towards smaller sizes at faint magnitudes can be clearly seen.

\subsection{Defining the S/G Separation Criterion}

\begin{figure*}
  \begin{center}
  \includegraphics{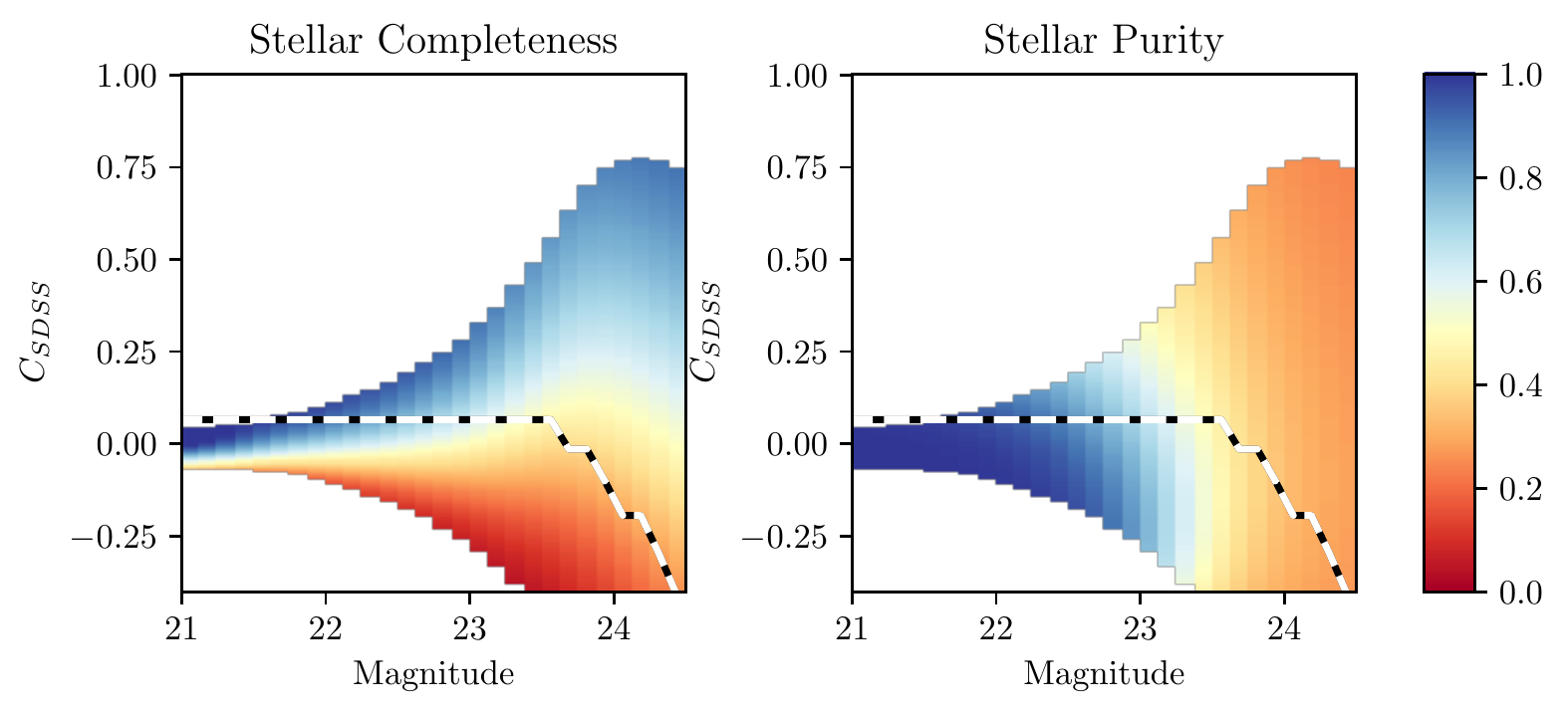}
\end{center}
  \caption{Stellar completeness (left) and purity (right) for an example model
  ($m_5 = 24.5$ and seeing FWHM of $0.7\arcsec$). At bright magnitudes, the
  completeness rapidly transitions from 0\% to 100\%
  as one moves vertically across the stellar locus at $C_\textrm{SDSS} = 0$, and the loss
  of purity (alternatively, increasing contamination) does not begin until very
  high $C_\textrm{SDSS}$ values. At faint magnitudes, the stellar locus is more diffuse
  and the galaxies begin to dominate in number at small sizes, causing a
  reduction in stellar purity for separator values close to the stellar locus.
  Our fiducial separator metric is shown by the dashed black-white line, which traces the point
    at each magnitude where the completeness is equal to the purity. Areas of very low stellar density
    are omitted from the colored shading.
  \label{fig:fisher_plots_pc}}
\end{figure*}

The result of this model is a map of the observed sizes and magnitudes of a set
of stars and galaxies in a given set of observations. In order to enable quantitative
discussion of S/G classification, hereafter we adopt $C_\textrm{SDSS}$ classifier. We have yet to apply any
sort of classification to the observed set of objects though, and there are many
issues that now arise when trying to do so. Defining a morphological classifier
is equivalent to drawing some line through the size-magnitude plane that defines
two regions (we will focus on a binary classification at the moment and defer
discussion of probabilistic classifiers to \S\ref{sec:probSG}). How one defines this line is entirely
dependent on the scientific goals one is trying to achieve. Different levels of
contamination and completeness may be acceptable to different science programs
and thus a binary classifier that is universally applicable cannot be uniquely defined.
We therefore define here a classifier which is sufficiently
\textit{representative} of S/G performance, but is not necessarily what one
would always use in practice. We draw a line in size-magnitude space such
that, at any given magnitude, the purity of the stellar sample (defined as
the number of correctly-labeled true stars divided by all objects classified as stars) is
equal to the completeness of the sample of stars (correctly-labeled true stars
divided by all true stars).

This separator is illustrated in Figure~\ref{fig:fisher_plots_pc}. For a given
magnitude column, these plots show what the purity and completeness of a stellar
sample would be if everything lower than a chosen y-axis point was classified as
a star. This choice of the classifier cutoff can vary with magnitude. As our
desired classification goal is for completeness to equal purity, we draw our
cutoff (dashed black-white line in Figure~\ref{fig:fisher_plots_pc}) such that for every
magnitude, the cutoff sits on the same color in both left and right panels of
the plot. This is equivalent to a diagonal line in the ROC curve plots. One can
see that at faint magnitudes, the classifier must be exceedingly stringent to
maintain this criterion, to the extent of reaching 30\% completeness or less
even by selecting only objects that appear smaller than the PSF (purely for
statistical reasons). At some point we decide that such a diminished sample of
stars is no longer useful, and we thus define a fiducial completeness level and
define the magnitude at which this level is reached as the S/G separation
``limit'' for this set of observations. This choice is again arbitrary in detail,
but will be useful for characterizing the relative S/G performance between
different observations. These choices act as means of reducing the
dimensionality of the model output, from the completeness and purity planes down
to a single number.

\subsection{Validating the Model with Stripe 82}

\begin{figure}
  \includegraphics{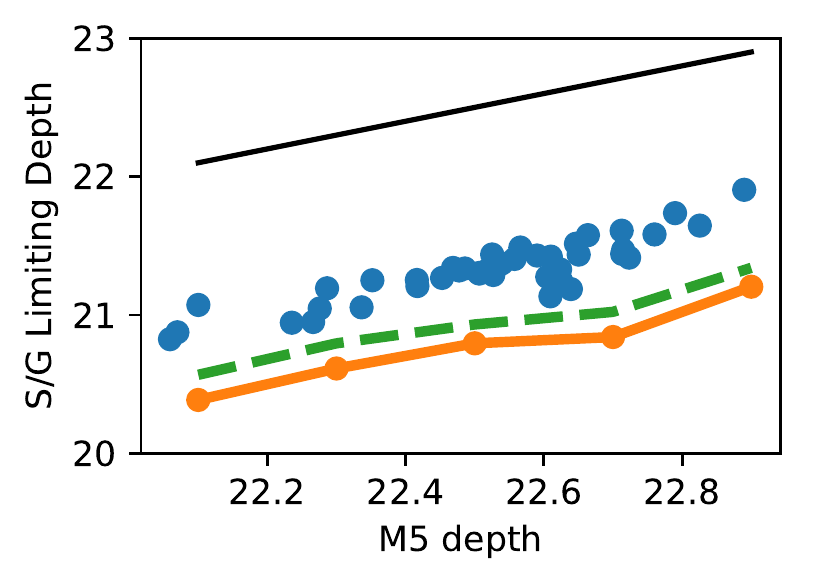}
  \caption{The blue points indiciate measured star/galaxy performance for several Stripe 82 runs (measured at a
  fiducial point where completeness is equal to purity at 80\%) compared to the
  $5\sigma$ limiting depth of each run. The orange dash-dotted line shows the
    model prediction for S/G performance over this range of depths (assuming Gaussian PSFs), while the dashed green
  line accounts for the overestimation of S/G performance due to the imperfect
  ground-based ``truth'' catalog. The black line shows where S/G performance
  equals $5\sigma$ depth, which is unrealistic in practice but shown as a visual
  aid.
  \label{fig:stripe82_fisher}}
\end{figure}

Before any extensive usage of this model, we need to verify that its
results accurately characterize the S/G separation performance of real
observations. To do so, we need a set of observations under varying seeing and
depth conditions, along with a set of ``truth'' labels for objects in these
observations. SDSS Stripe 82 fits these criteria as it consists of numerous
repeat observations of the same equatorial stripe, yielding approximately 80
measurements
of any given patch of sky, and which were observed in both
photometric and various degraded observing conditions \citep{abazajian09}.
We compare the S/G measurements from these observations those from the Dark
Energy Survey, Data Release 1 \citep{abbott18}, which is significantly deeper
and has better seeing than the SDSS data. In the r-band, the DES data reaches a
$\textrm{SNR}=10$ at $24.08$, and has a median seeing of $0.96\arcsec$.
This extra depth
enables us to use this deep catalog as an approximate ``truth'' table, since objects which
are on the verge of being resolved in the single epoch data, and thus which
contribute most to any changes in S/G performance, will be identified
in the coadd data. There will still be some unresolved galaxies in this coadded
data, and we will discuss the implications and effects of this contamination
below.

To model these Stripe 82 performance, we created a stellar density distribution that was
normalized to the galaxy density at an r-band magnitude of 20.8, as is seen in
the observed Stripe 82 coadd data. The galaxy distribution was the same as
measured in COSMOS (see Section~\ref{sec:COSMOS}). The stellar density model was
slightly rising, such that it doubled after two magnitudes
of increased depth, again to approximate the observed density distribution.

While our model allows us to specify the observed depth, often characterized by
the $5\sigma$ limiting magnitude or ``$m_5$'', separately from the seeing, in
practice these variables are strongly correlated for data from a single
telescope even under varying observing conditions. For this reason, we test our model
on SDSS assuming that only the $m_5$ depth varies independently, and the seeing
is linked to this by $m_5 = \log(0.7/\theta) + C$, where $\theta$ is the seeing
FWHM and C is a constant fit to the SDSS data \citep{2019ApJ...873..111I}. Reducing the
model to one parameter simplifies validation and visualization, but this
restriction will be lifted when using the model to make predictions.

For computing the S/G limiting magnitude of the individual Stripe 82 runs, we
follow a similar procedure as in the modeling to define the value of $C_\textrm{SDSS}$
for each magnitude bin at which the stellar completeness is equal to the purity.
This takes advantage of the deeper data in measuring completeness and purity,
which most surveys normally lack (otherwise they would simply use the deeper
data), but does not introduce the extra uncertainty of trying to
estimate these parameters from only the shallow data. This additional
information does not improve S/G separation, it only improves our measurement of
S/G performance.

Figure~\ref{fig:stripe82_fisher} shows the result of this verification exercise.
The blue points show the depth of individual Stripe 82 runs, while the solid orange
line shows the depth estimated using our modeling. In general the reported
Stripe 82 S/G depth is somewhat deeper that predicted by the modeling, but some
of this difference certainly comes from the ground-based reference catalog used
for the comparison, which itself has some contamination fraction. Thus the
measured Stripe 82 points slightly \textit{overestimate} the S/G depth, by
failing to recognize some unresolved objects as contaminant galaxies. We roughly
estimate the significance of this effect by drawing the dashed line in
Figure~\ref{fig:stripe82_fisher}, which mimics this unrecognized contamination
by modeling a less-stringent S/G separation purity, equivalent to a 20\%
unrecognized contamination rate. The resulting model tracks the change of S/G
performance with observation depth quite well, although there remains a modest
offset of $\sim 0.3$ mag between the absolute predicted performance and the measured performance.
As our main interest is in using the model for understanding the seeing and
depth dependence of S/G performance, we consider this level of agreement
acceptable.

\subsection{Model Lessons}

\begin{figure}
  \plotone{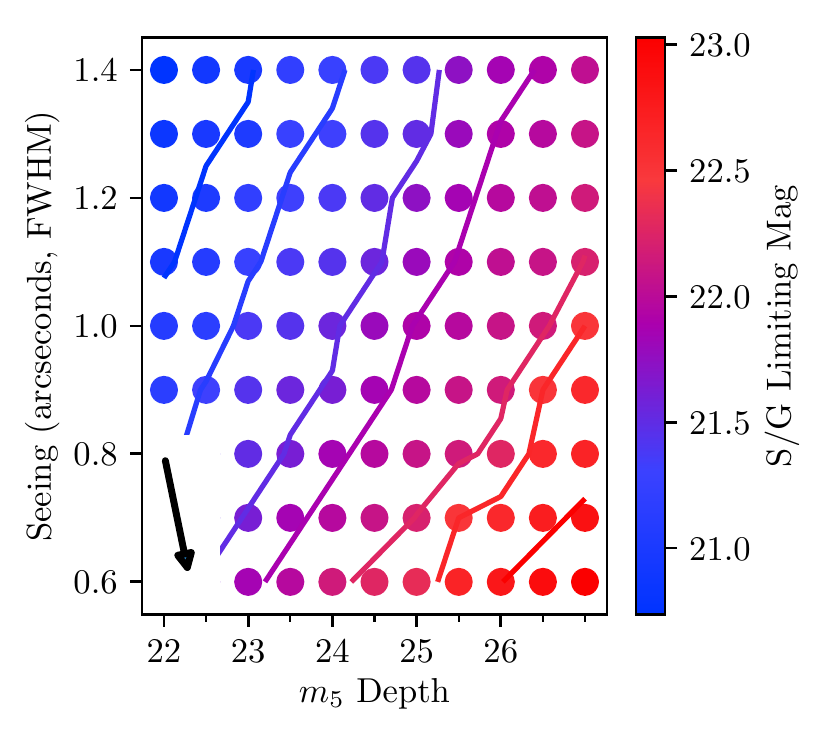}
  \caption{S/G separation limit, defined as completeness equal to purity reaches
  80\%, as a function of survey seeing and $5\sigma$ limiting depth. The
  bottom left arrow indicates the vector of changing observational seeing;
  i.e., the difference between the head and tail represents the $m_5$
  improvement resulting from a change in seeing alone.
  \label{fig:sglimit_model_grid}}
\end{figure}

\begin{figure}
  \plotone{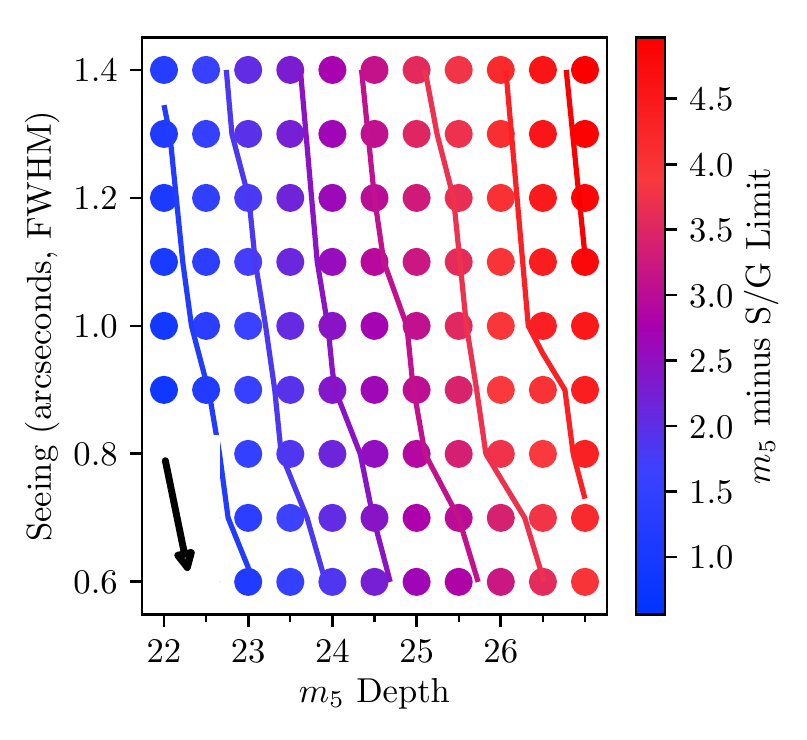}
  \caption{Same as Figure~\ref{fig:sglimit_model_grid}, but S/G limiting magnitudes have had the
  $5\sigma$ depth subtracted. This figure thus shows how the S/G limit
  diverges from the photometric depth at faint magnitudes.
  \label{fig:sg_rel_model_grid}}
\end{figure}

\begin{figure*}
  \begin{center}
  \includegraphics{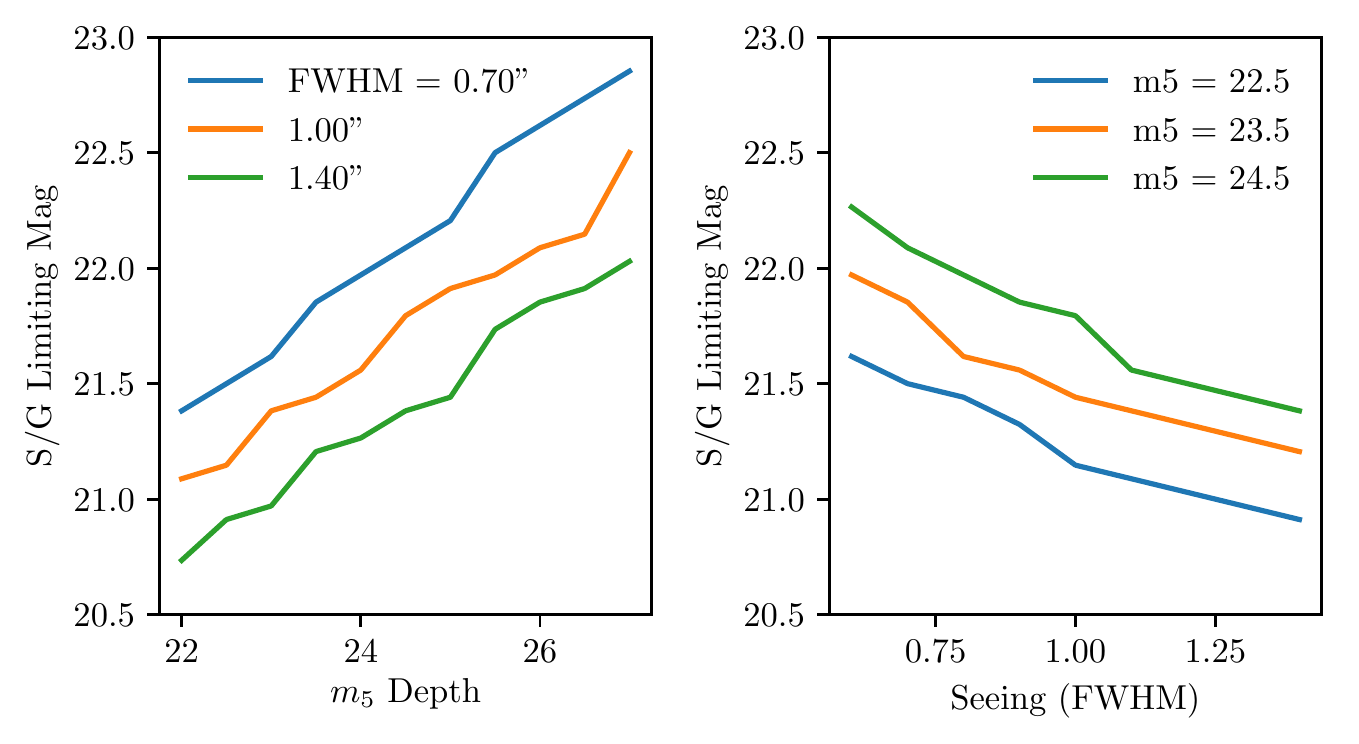}
  \end{center}
  \caption{Left: predicted S/G limiting depth for three example seeing
  values. These correspond to horizontal slices of
  Figure~\ref{fig:sglimit_model_grid}, and have an average slope of $0.27$ mag  S/G per mag $m_5$. Right:
  predicted S/G depth for example $m5$ depths, corresponding to
  \textit{vertical} slices of Figure~\ref{fig:sglimit_model_grid}. The average slope is $-1.0$ mag S/G per arcsecond of seeing.
    \label{fig:model_predictions_m5}}
\end{figure*}

For the purpose of discussion in this section, we adopt as the S/G limiting depth
the magnitude at which both purity and completeness for stellar sample are 80\%.
We evaluate it over a grid of $5\sigma$ limiting depths and seeing values, as
illustrated in Figures~\ref{fig:sglimit_model_grid} and \ref{fig:sg_rel_model_grid}.
Several qualitative and quantitative conclusions can be readily derived from
features visible in these figures.

These figures show that at constant seeing the S/G limiting depth does not improve
as fast as $5\sigma$ limiting depth. For example, in 1 arcsec seeing, the difference
$m_5-m_\textrm{SG}$ increases from about 1.5 mag at $m_5=22.5$ to about 2.5 mag at
$m_5=24.5$. In other words, although $m_5$ improved by 2 magnitudes, $m_\textrm{SG}$
improved by only 1 magnitude. The same conclusion is valid for other values of
seeing (the lines of constant $m_5-m_\textrm{SG}$ difference in Figure \ref{fig:sg_rel_model_grid}
are nearly straight and parallel to each other), and it is a direct result of increasing
galaxy-to-star count ratio and decreasing intrinsic galaxy size with magnitude.

When seeing varies, its impact on $m_5$ has to be taken into account via $\Delta m_5
= 2.5\log_{10}(\theta_2/\theta_1)$; for example, when seeing improves from 1.4 arcsec
(typical for SDSS) to 0.7 arcsec (anticipated as typical for LSST), $m_5$ improves
by 0.75 magnitudes. In this case, figures show that  $m_5-m_\textrm{SG}$ difference stays
approximately constant and thus $m_\textrm{SG}$ improves by about 0.75 magnitudes.
In order to improve $m_\textrm{SG}$ by the same amount in constant seeing, $m_5$
would have to be improved by at least 0.75 magnitudes, which implies an
increase of exposure time (assuming that background brightness and other
observing properties remain unchanged) of at least a factor of 4.

Finally, it is illustrative to compare the performance of star-galaxy separation
in our model for fiducial seeing and $m_5$ corresponding to SDSS (seeing of
1.4 arcsec and $m_5=22.5$) and LSST (seeing of 0.7 arcsec and $m_5=27$).
While the LSST performance, relative to that of SDSS, will be affected negatively
by the increasing galaxy-to-star count ratio and decreasing intrinsic galaxy size
with magnitude, it will benefit from better seeing and deeper data. As
Figures~\ref{fig:sglimit_model_grid} and \ref{fig:sg_rel_model_grid} show,
if LSST had the same seeing as SDSS, due to $m_5$ improvement of 4.5 magnitudes,
$m_\textrm{SG}$ would improve by (only) 1.5 mag (from about 21.0 to 22.5). However,
because the seeing is also improved, $m_\textrm{SG}$ would improve by about 2.0-2.5
magnitudes (to 23.0-23.5). However, note that this statement is valid only 
for the definition of $m_\textrm{SG}$ adopted here (and it depends strongly on the 
actual star-to-galaxy count ratio). 

\section{Probabilistic S/G Separation Cookbook}
\label{sec:probSG}

While the preceding sections have focused on the measurement of images and the
expected performance of a classifier, they have not addressed how to best use
the results from a catalog of measurements. In this section we present a brief
``cookbook'' for how a measurement such as $C_\textrm{SDSS}$ can be used to construct a
probabilistic S/G separator, which will enable the combination of information
from multiple measurements in a theoretically sound manner.

The basic outline of the process is as follows:

\begin{enumerate}

  \item \textbf{Obtain both survey data and a set of accurate labels.}
  The most common method for obtaining star and galaxy labels is with
  space-based observations, where galaxies are readily resolved, but other
  methods such as spectroscopy could also provide this classification. We will
  assume that these labels are perfectly accurate in our description below.
  It is also important that data used for training spans the range of SNR and seeing
  conditions present in the survey for which the classifier will be used, as the
  probabilistic classification is dependent on the noise properties of the
  training set matching that of the target observations.

  \item \textbf{Compute} $p(c|m, \theta_\textrm{psf}, \textrm{SNR}, [S/G])$. That is, we
  must construct a function that transforms the raw measurement $c$ from the classifier algorithm (e.g.
  the SDSS model minus PSF magnitude, and which can have arbitrary scaling) into a properly
        normalized \textit{probability} $p(c|\dots)$. One simple way to do
  this, ignoring for the moment the dependence on magnitude, SNR, and seeing, is to
  use the high resolution data to separate stars from galaxies, then for each
  set compute a kernel density estimator on the classifier values measured in
  the target survey data (one can think of this as a slightly more sophisticated
  version of histogramming the data as a function of $c$). 
  This function must then be normalized such that the
  integral over the classifier value $c$, at a given magnitude, is unity. This is
  $p(c | m, [S/G], \dots)$. Note that this function encodes information about how the
  measurement of $c$ responds to intrinsic object sizes and shapes in the
  presence of measurement noise, along with the distribution of object shapes
  and sizes in the sample. Because of this complexity it is more effective to
  fit $p(c|m, \dots)$ empirically than to develop a forward model for this
  function. Additional parameters could be included here to handle effects such as, e.g., PSF
  chromaticity \citep{carlsten18}.

  \item \textbf{Combine multiple measurements of $p(c|S, \dots)$.} For objects that are
  measured multiple times, either in different images or different filters, the
  appropriate way to combine these measurements is by computing $p(c|S, \dots)$ for
  each of the different measurements individually and multiply these factors
  together, i.e.,
  \eq{
         L_S = \prod_{i=1}^N \,  p(c_i|m,\theta_\textrm{psf},{\rm SNR},S),
  }
  and analogously for $L_G$. This cannot be done with the classifier value $c$ directly, which
  is why step 2 in our outline is critical. While it is important that all individual probabilities are
  properly normalized (the integral over $c$ must be unity), the combined data
  probability need not be a proper PDF since we will only use it in the ratio of
  $L_S/L_G$, where the normalization cancels.

  \item \textbf{Use Bayes' theorem to obtain $p(G|c, \dots)$ and $p(S|c, \dots)$.}
  Because we require that $p(S | \dots) + P(G | \dots) = 1$, we can use Bayes'
  theorem to obtain
  \eq{\label{eq:combSG} 
     p(G|\{c_i\},m,\theta_\textrm{psf},{\rm SNR}) = \\
        \left[1 + \frac{L_S\,p(S|m)}{L_G\,p(G|m)}\right]^{-1}.
  }
  This equation combines all of the measurements of an object with a prior on
  the ratio of stars to galaxies. This choice of prior is extremely important,
  as the ratio of stars to galaxies can vary by more than an order of magnitude
  across the sky. While one could have empirically used the fitting of the
  training sample to estimate $p(S)$ or $p(G)$ directly, it would have then
  carried an assumption that the relative number density of stars and galaxies
  is the same in the training sample as in the target sample of interest.
  Computing $p(c|[S/G])$ first decouples these two samples, enabling a training
  sample at high Galactic latitude, for example, to be used for calibrating S/G
  separation of a survey at a wide range of stellar densities (though the
  effects of crowding will at some point alter the properties of the
  classifier measurement).

\end{enumerate}

The resulting probabilities $p(G|\dots)$ and $p(S|\dots)$ can be directly used in
analysis, or as an input to a judiciously chosen classification procedure. For
example, if the use case needs a very complete sample of stars, then all
objects with $p(S|\dots)>0.5$ could be classified as stars, but if a very clean
sample of stars is required, then $p(S|\dots)>0.99$ might be a more appropriate
condition. As before, there is no universal optimum and the choice of a position
on ROC curve depends on the chosen completeness vs. purity tradeoff.
Alternatively, if the desired scientific quantity is the number of stars or
galaxies in a region of sky or other parameter-space, simply summing the $p(S)$
or $p(G)$ over all objects in the target region produces an estimate of the
number in either class.

\begin{figure}
  \plotone{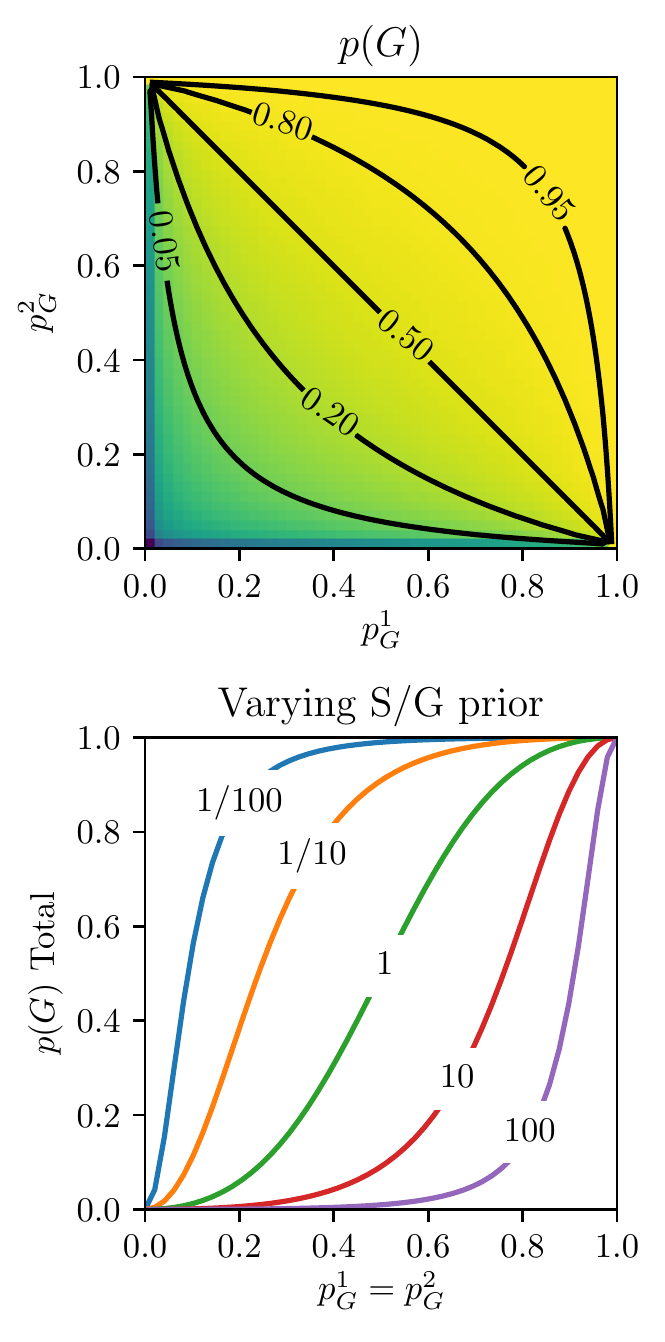}
  \caption{\textit{Top}: Combined probabilitiy of that an observed object is
  a galaxy, given two input measurements $p_G^1$ and $p_G^2$, and assuming an
  equal density of stars and galaxies as a prior. \textit{Bottom}: Combined
  probability, given two input measurements which are equal to each other,
  but under varying assumed priors for the ratio of stars to galaxies. These correspond to slices
    along the lower-left to upper-right diagonal of the top panel.
  \label{fig:combined_measurements}}
\end{figure}

As an illustration of probabilistic combination of multiple measurements using
eq.~\ref{eq:combSG}, we consider a simple case of just two measurements (note 
that they correspond to likelihoods), 
\eq{
        p_G^i \equiv p(c_i|m,\theta_\textrm{psf},{\rm SNR},G),
}
with $i=1,2$ and $p_S^i = 1 - p_G^i$. These two measurements could be based
on two images in the same bandpass, two images in different bandpasses, or 
perhaps correspond to one morphological and one color-based measurement. 
It follows from eq.~\ref{eq:combSG} that the final probability that the source
under consideration is resolved is 
\eq{
    p(G) = \frac{1}{\displaystyle 1 + \frac{p(S|m)}{p(G|m)} \frac{(1-p_G^1)(1-p_G^2)}{p_G^1 \, p_G^2} }.
}
Note that $p(G)=0$ when at least one of $p_G^1$ and $p_G^2$ is zero, and
$p(G)=1$ when at least one of $p_G^1$ and $p_G^2$ is unity, as intuitively
expected. For a given value
of the prior $p(S|m)/p(G|m)$, $p(G)$ is a two-dimensional function of $p_G^1$ and 
$p_G^2$, illustrated in Figure~\ref{fig:combined_measurements}. The figure shows
how high confidence measurements ($>0.9$) can outweigh ambiguous
measurements, leading to high confidence of the resulting classification.

Further insight can be obtained by taking the $p_G^1=p_G^2$ slice through
the top panel of Figure~\ref{fig:combined_measurements}.
With abbreviations $p_\textrm{SG} \equiv p(S|m)/p(G|m)$ and $x \equiv p_G^1=p_G^2$, 
\eq{
                 p(G) = \frac{x^2}{x^2 + p_\textrm{SG}(1-x)^2}, 
}
which is illustrated in the lower panel of
Figure~\ref{fig:combined_measurements}. As evident, $p(G)$ is a monotonic
function of $x$,
with a value of $x$ that corresponds to $p(G)=0.5$ strongly dependent on 
the prior $p_\textrm{SG}$. Consider two uninformative measurements, $x=0.5$. When
$p_\textrm{SG}=1$, the final probability remains uninformative, $p(G)=0.5$. However,
when $p_\textrm{SG}=10$, for example, this prior that strongly favors $S$ results in 
$p(G)=0.09$ (and symmetrically, $p(G)=0.91$ for $p_\textrm{SG}=0.1$). On the other 
hand, when measurements strongly favor $G$, e.g. $x=0.9$, then $p(G)=0.99$
for $p_\textrm{SG}=1$ and $p(G)=0.89$ even when $p_\textrm{SG}=10$ (and $p(G)=0.999$ for 
$p_\textrm{SG}=0.1$). Overall, ambiguous measurements lean towards the prior, while
high confidence measurements are required when the underlying prior strongly
disfavors a particular classification. As always, an evaulation of the ROC
curve is still required to deliver the desired completeness and contamination
properties for a given scientific use case.

\section{Discussion and Conclusion}
\label{sec:conclusions}

Our work has also shown that many of the commonly used measurement techniques
for S/G separation are all closely related to each other, and also related to
the theoretically optimal technique described by \citet{Sebok}. The resulting
performance of these classifiers is thus very similar, with the primary
differences resulting from their treatment of noise and the evolution of their
numerical values with SNR or depth. These similarities suggest that the
measurements on the pixels themselves are unlikely to see dramatic improvement
from new algorithms. There are of course simplifying assumptions in our analysis
that may be addressed by practical implementations of these techniques, and
gains in S/G performance to be had from such methods, but the basic comparison
of a broadened profile with a PSF profile appears firmly planted.

Our modeling of the theoretical performance of an idealized S/G classifier
emphasizes the importance of the image SNR on the resulting completeness and
contamination. This is often under-appreciated, and angular resolution is often
assumed to be the key factor in S/G performance. The distinction between these two
mechanisms is subtle but significant. Low SNR objects have
poorly-constrained size measurements, making it difficult to distinguish
PSF-shaped objects from broadened ones. Deeper observations can increase the SNR of
objects, which enables more precise shape measurement and thus better S/G
separation. Such deeper observations can come from longer exposure times, better
seeing, or by observing on a larger-aperture telescope.

Improved seeing also increases the observed size difference between point
sources and galaxies of a given size, and hence enables less precise (lower
SNR) size measurements to successfully distinguish stars from galaxies. This
effect is obviously well-known, but we emphasize in this work that a substantial
portion of its apparent effectiveness is due to the improved SNR in better seeing conditions.

To fully take advantage of the information in high SNR images, however,
the image PSF must be precisely characterized. It is beyond the scope of this
work to quantify the effect of errors in the PSF model, but it is clear that
systematic errors in S/G separation can be introduced by the use of a poor
quality PSF model. Extra caution must be used when trying to classify small
objects (relative to the PSF size) using high SNR data, since spatial or
chromatic PSF variations or detector effects may be relatively more significant,
rather than obscured in the noise.

Surveys on 8-m class telescopes, such as the Hyper Suprime-cam Survey and LSST,
will place strong demands on S/G separation, relying on SNR to overcome the
increasing numbers of galaxies at faint magnitudes. Extracting the most stellar
and Galactic science from these surveys will require careful attention at all
stages of survey design, image processing, and statistical treatment of the
resulting catalogs.

Additionally, the challenge of deep S/G separation on 8-meter class surveys will
increase the importance of combining all available information when classifying
objects, across images, passbands, and including non-morphological information
such as colors or proper motion measurements. We have outlined a blueprint of a
procedure for this. Converting each individual type of measurement to a
probabilistic form also enables the user to apply appropriate priors, such as
models of the star and galaxy density distributions or measurements from other
surveys. Placing S/G separation on a rigorously probabilistic basis will
maximize the scientific return of these surveys.

\acknowledgements

The authors thank Michael Wood-Vasey, Sophie Reed, and the anonymous referee for their helpful
comments which improved the work, and Peter Yoachim for work on the galaxy count and size model.

This material is based upon work supported by the National Science Foundation
under Cooperative Agreement 1258333 managed by the Association of
Universities for Research in Astronomy (AURA), and the Department of Energy
under Contract No. DE-AC02-76SF00515 with the SLAC National Accelerator
Laboratory. Additional LSST funding comes from private donations, grants to
universities, and in-kind support from LSSTC Institutional Members.

\appendix
\section{Fisher Information calculation}
\label{appendix}

In this appendix we derive the Fisher information for a set of pixels $f_n$
which are drawn from a Gaussian distribution, and where the expected mean values are
\begin{equation}
  f_n = C_\textrm{gal} g_i(\theta) + B,
\end{equation}
where $g_i(\theta)$ is the galaxy model with a vector of shape parameters
$\theta$, B is a constant background level across all pixels, and $C_\textrm{gal}$ is the
total flux of the object.

The Fisher information is defined as
\begin{equation}
\mathcal{I}_\theta = \mathbb{E}_{f_1, ..., f_n \sim f_\theta^M}\left[
\left( \frac{\partial}{\partial \theta
} \ln L(f_1, ..., f_n; \theta) \right)^2
\right]
\end{equation}

where $f_n$ is the measured value of pixel $n$. Our likelihood function $L(f_1, ...,
f_n; \theta)$ was defined earlier in Equation 3 as $p(D|M, C, \theta)$.
Inserting this likelihood function and dropping terms inside the partial
derivative with no dependence on $\theta$ yields
\begin{equation}
\mathcal{I}_\theta = \mathbb{E}_{f_1, ..., f_n \sim f_\theta^M}\left[
\left( \frac{\partial}{\partial \theta}
\sum_{i=0}^{N} \frac{-[f_i - F_i(\theta) ]^2}{2\sigma_i^2}
\right)^2
\right],
\end{equation}
where $F_i(\theta)$ denotes the expected value for pixel $i$ of the model being
fit to the observations (i.e., the noise-free version of Equation~\ref{eq:fG}.)

Evaluating the derivative,
\begin{equation}
\mathcal{I}_\theta = \mathbb{E}_{f_1, ..., f_n \sim f_\theta^M}\left[
\left(
    \sum_{i=0}^{N}
    -
    \frac{\partial F_i(\theta)}{\partial \theta}
    \frac{\left[f_i - F_i(\theta)\right]}{\sigma_i^2}
    \right)^2 \right].
\end{equation}

We then assume that the measurement residuals are uncorrelated, that is,
\begin{equation}
\mathbb{E}_{f_1, ..., f_n \sim f_\theta^M}\left[
\left[f_i - F_i(\theta)\right]
\left[f_j - F_j(\theta)\right]
\right] = 0
\end{equation}
for all $(i,j)$ where $i\ne j$. This enables us to obtain
\begin{equation}
\mathcal{I}_\theta =
\mathbb{E}_{f_1, ..., f_n \sim f_\theta^M}\left[
    \sum_{i=0}^{N}
    \left( \frac{1}{\sigma_i^2} \frac{\partial F_i(\theta)}{\partial \theta}
    \right)^2
    \left[f_i - F_i(\theta)\right]^2 \right].
\end{equation}

Pulling the summation out of the expectation value, and using the fact that
\begin{equation}
\mathbb{E}_{f_i \sim f_\theta^M}\left[
(f_i - F_i(\theta))^2 \right] = \sigma_i^2
\end{equation}
we obtain
\begin{equation}
    \mathcal{I}_\theta = \sum_{i=0}^{N}
    \frac{1}{\sigma_i^2}
    \left(
    \frac{\partial F_i(\theta)}{\partial \theta}
    \right)^2.
\end{equation}

This is directly analogous to the solution in the case of Poisson noise (from
\citealt{mendez13}), which is
\begin{equation}
  \mathcal{I}_\theta = \sum^N_{i=1} \frac{1}{\lambda_i(\theta)}
  \left( \frac{d\lambda_i(\theta)}{d\theta} \right)^2
\end{equation}
where $\lambda_i$ is the expected value in pixel $i$.

\bibliographystyle{apj}
\bibliography{paper}

\end{document}